\documentclass[preprint,11pt]{elsarticle}

\usepackage{amsthm,amsmath}
\usepackage{amsfonts}       
\usepackage{booktabs}       
\usepackage{graphicx}
\usepackage{placeins} 		
\usepackage{adjustbox}
\usepackage{caption}
\usepackage[margin=1in]{geometry}
\usepackage{bm}
\usepackage{csvsimple}
\usepackage{siunitx}
\usepackage{tabularx}
\usepackage{fp}
\usepackage{bm}
\usepackage[hyphens]{url}
\newcommand{\beginsupplement}{%
        \setcounter{table}{0}
        \renewcommand{\thetable}{S\arabic{table}}%
        \setcounter{figure}{0}
        \renewcommand{\thefigure}{S\arabic{figure}}%
     }
\newcommand\BoldCaseThreeAll[1]{\FPiflt{#1}{4.5}\num[math-rm=\mathbf]{#1}\else\num{#1}\fi}
\newcommand\BoldCaseTwoAll[1]{\FPiflt{#1}{12}\num[math-rm=\mathbf]{#1}\else\num{#1}\fi}
\newcommand\BoldCaseOneAll[1]{\FPiflt{#1}{11}\num[math-rm=\mathbf]{#1}\else\num{#1}\fi}

\newcommand\BoldCaseThreeSparse[1]{\FPiflt{#1}{2.5}\num[math-rm=\mathbf]{#1}\else\num{#1}\fi}
\newcommand\BoldCaseTwoSparse[1]{\FPiflt{#1}{4}\num[math-rm=\mathbf]{#1}\else\num{#1}\fi}
\newcommand\BoldCaseOneSparse[1]{\FPiflt{#1}{6}\num[math-rm=\mathbf]{#1}\else\num{#1}\fi}

\newtheorem{theorem}{Theorem}
\newtheorem{prop}{Proposition}

\journal{Computational Statistics and Data Analysis}
\begin{document}

\begin{frontmatter}
\title{Numerical Characterization of Support Recovery in Sparse Regression with Correlated Design}

\author{$\text{Ankit Kumar}^{\text{\emph{a,b,c}}}$}
\author{$\text{Sharmodeep Bhattacharyya}^{\text{\emph{f}}}$}
\author{$\text{Kristofer Bouchard}^{\text{\emph{b,c,d,e,*}}}\corref{cor1}$}
\ead{kebouchard@lbl.gov}
\cortext[cor1]{Corresponding author}

\address[1]{Deperatment of Physics, University of California, Berkeley, Berkeley, CA 94720}
\address[2]{Redwood Center for Theoretical Neuroscience, University of California, Berkeley, Berkeley, CA 94720}
\address[3]{Biological Systems and Engineering Division, Lawrence Berkeley National Laboratory, Berkeley, CA 94720}
\address[4]{Computational Research Division, Lawrence Berkeley National Laboratory, Berkeley, CA 94720}
\address[5]{Helen Wills Neuroscience Institute, University of California, Berkeley, Berkeley, CA 94720}
\address[6]{Department of Statistics, Oregon State University, Corvallis, Oregon 97331}

\begin{abstract}
    Sparse regression is frequently employed in diverse scientific settings as a feature selection method. A pervasive aspect of scientific data that hampers both feature selection and estimation is the presence of strong correlations between predictive features. These fundamental issues are often not appreciated by practitioners, and jeapordize conclusions drawn from estimated models. On the other hand, theoretical results on sparsity-inducing regularized regression such as the Lasso have largely addressed conditions for selection consistency via asymptotics, and disregard the problem of model selection, whereby regularization parameters are chosen. In this numerical study, we address these issues through exhaustive characterization of the performance of several regression estimators, coupled with a range of model selection strategies. These estimators and selection criteria were examined across correlated regression problems with varying degrees of signal to noise, distribution of the non-zero model coefficients, and model sparsity. Our results reveal a fundamental tradeoff between false positive and false negative control in all regression estimators and model selection criteria examined. Additionally, we are able to numerically explore a transition point modulated by the signal-to-noise ratio and spectral properties of the design covariance matrix at which the selection accuracy of all considered algorithms degrades. Overall, we find that SCAD coupled with BIC or empirical Bayes model selection performs the best feature selection across the regression problems considered. 
\end{abstract}



\end{frontmatter}
\section{Introduction}

In the last several decades, significant research in the mathematics and statistics communities has been directed at the problem of reconstructing a $k$-sparse vector from noisy, linear observations. In its simplest form, one is concerned with inference within the following model:

\begin{align}
    \bm{y} = \bm{X} \beta + \epsilon
    \label{linmodel}
\end{align}

with $\bm{y} \in \mathbb{R}^n, \bm{X} \in \mathbb{R}^{n \times p}$ and $\beta \in \mathbb{R}^p$ is a k-sparse vector. The noise is i.i.d, $\epsilon \in \mathbb{R}^n, \epsilon_i \sim \mathcal{N}(0, \sigma^2)$, and the observational model is Gaussian, $y_i \sim \mathcal{N}(\bm{X}_i \beta, \epsilon_i)$. The sparse linear model is employed in diverse scientific fields \cite{satija_spatial_2015, waldmann_evaluation_2013, tibshirani_lasso_1997, steyerberg_towards_2014, wright_sparse_2010}. In real world applications, it is also commonly the case that the design or covariate matrix X is correlated, so that the columns of $\bm{X}$ can not be taken to be i.i.d. In this setting, the correct identification of non-zero elements of $\beta$, which is crucial for scientific interpretability, is especially challenging. Yet, a systematic exploration of the effect of correlations between the covariates on the recoverability of $\beta$ is lacking. 

Statistically optimal sparse estimates of $\beta$ within (\ref{linmodel}) are returned by the solution to the following constrained optimization problem:

\begin{align}
    &\min ||y - \bm{X} \beta||_2^2 \\
    &||\beta||_0 \leq \lambda
    \label{l0}
\end{align}

Finding the global minima of problem (\ref{l0}) is NP-hard, though recent progress has been made in computationally tractable approaches \cite{zhu_polynomial_2020, bertsimas_best_2016}. The most common approach is to relax the $l_0$ regularization. In this work, we focus on the Lasso, Elastic Net, SCAD, MCP \cite{tibshirani_regression_1996, zou_regularization_2005, zhang_nearly_2010, fan_variable_2001}, and $\text{UoI}_{\text{Lasso}}$, an inference framework we introduced in \cite{bouchard_union_2017} that combines stability selection and bagging approaches to produce low variance and nearly unbiased estimates. To select the regularization strength or otherwise compare between candidate models returned between these estimators, one must employ a model selection criteria such as cross-validation or BIC. While the literature on sparsity inducing estimators and model selection criteria is vast, studies that consider the interaction of particular choices of estimator and model selection criteria are lacking. In particular, no systematic exploration of the impact of choice of estimator \emph{and} model selection criteria on the selection accuracy of the resulting procedure when the predictive features exhibit correlations has been carried out. In this work, we address this gap by performing systematic numerical investigations of the selection accuracy performance of several estimators and model selection criteria across a broad range of regression designs, including diverse correlated design matrices. Section 2 summarizes prior theoretical and empirical work on model selection and compressed sensing. We also discus a scalar parameterization of signal strength in correlated sparse regression borrowed from \cite{m._j._wainwright_information-theoretic_2009} that we call $\alpha$. In section 3, we outline the scope of this study and the evaluation criteria used. In Section 4 we present the main results. We characterize the impact of correlated design on the false negative and false positive discovery rates, as well as the magnitude of coefficients likely to be falsely set to zero or false assigned non-zero values. We reveal that estimators and selection methods display a remarkable degree of universality with respect to the correlation strength (quantified by $\alpha$). We also identify the best performing combinations of estimator and selection methods under various signal conditions. Connections to prior theoretical work and concrete recommendations for practitioners are provided in Section 5. 

\section{Review of Prior Work}

The statistical theory of the sparse estimators considered in this work is vast and we do not attempt to review it all here. Our particular focus is on characterizing finite sample selection accuracy, especially in the context of correlated design. The asymptotic oracular selection performance of the SCAD and MCP are well known \cite{fan_variable_2001, zhang_nearly_2010} and require only mild conditions on the design matrix. For the Lasso, one must impose an irrepresentible condition to guarantee asymptotic selection consistency \cite{zhao_model_2006}. The finite sample implications of these differing requirements have not been explored. A series of works have addressed the correlated design problem by devising regularizations that tend to assign correlated covariates similar model coefficients \cite{li_graph-based_2018, figueiredo_ordered_2016, buhlmann_correlated_2013, bogdan_statistical_2013, tibshirani_sparsity_2005, witten_cluster_2014}. In fact, the Elastic Net was the first estimator introduced to exhibit this type of ``grouping'' effect \cite{zou_regularization_2005}. However, this type of behavior can be undesirable in many real data applications where covariates may be correlated, yet still contribute heterogenously to a response variable of interest.

When the true model generating the data is contained amongst the candidate model supports, the BIC and gMDL have asymptotic guarantees of selection consistency \cite{zhao_model_2006}. Extensions of these results to the high dimensional case are available \cite{kim_consistent_2012}, but fall outside the scope of this work. Implicit in these theoretical results is that one can evaluate the penalized likelihoods on all $2^p$ candidate model supports \cite{shao_asymptotic_1997}. Practically, one first assembles a much smaller set of candidate model supports using a regularized estimators. To this end, the use of the BIC with SCAD has been shown to be selection consistent \cite{wang_tuning_2007}.
\begin{table}[t!]
\hskip0.1cm\begin{tabular}{ll|}
\hline
\multicolumn{1}{|l@{\hspace{5.1em}}|}{Estimator}   & \multicolumn{1}{l@{\hspace{25.62em}}|}{Regularization}                                                                                                                                                \\ \hline
\multicolumn{1}{|l|}{Lasso}       & \multicolumn{1}{l|}{$\lambda |\beta|_1$}                                                                                                                                           \\ \hline
\multicolumn{1}{|l|}{Elastic Net} & \multicolumn{1}{l|}{$\lambda_1 |\beta|_1 + \lambda_2 |\beta|_2^2$}                                                                                                                  \\ \hline
\multicolumn{1}{|l|}{SCAD}        & \multicolumn{1}{l|}{$\int_0^{|\beta|} dx\left(\lambda \mathbb{I}(|\beta| \leq \lambda) + \frac{(\gamma \lambda -x)_+}{(\gamma - 1)\lambda} \mathbb{I}(|\beta| > \lambda) \right)$} \\ \hline
\multicolumn{1}{|l|}{MCP}         & \multicolumn{1}{l|}{$\int_0^{|\beta|} dx \left(1 - \frac{x}{\gamma \lambda} \right)_+$}                                                                                            \\ \hline
\multicolumn{1}{|l|}{$\text{UoI}_{\text{Lasso}}$}       & $\lambda|\beta|_1$ across bootstraps, see \cite{bouchard_union_2017}  \\ \hline 
\end{tabular}\\
\newline
\vspace*{1pt}
\newline
\resizebox{\columnwidth}{!}{ 
\begin{tabular}{|l|l|}
\hline
Model Selection Criteria &                                                                                                                                                                                                                                                                                                       \\ \hline
Cross-Validation         & $R^2$ averaged over 5 folds                                                                                                                                                                                                                                                                           \\ \hline
BIC                      & $2\log|y - X \hat{\beta}|_2^2 - \log(n) |\hat{\beta}|_0$                                                                                                                                                                                                                                              \\ \hline
AIC                      & $ 2\log|y - X \hat{\beta}|_2^2 - 2 |\hat{\beta}|_0$                                                                                                                                                                                                                                                   \\ \hline
gMDL \cite{hansen_model_2001}                     & $\begin{cases} \frac{\hat{k}}{2} \log \left( \frac{n - \hat{k}}{\hat{k}} \frac{y^\top y - |y - \hat{y}|_2^2}{|y - \hat{y}|_2^2}\right) + \log n & \text{if} \;\; R^2 > \frac{\hat{k}}{n} \\ \frac{n}{2} \log \left(\frac{y^\top y}{n} \right) + \frac{1}{2} \log(n) & \text{otherwise} \end{cases}$                           \\ \hline
Empirical Bayes \cite{george_calibration_2000}          & $ 2\log|y - X \hat{\beta}|_2^2 - \begin{cases} \hat{k} + \hat{k} \log(\hat{y}^\top \hat{y}) - \hat{k} - 2((p - \hat{k}) \log(p - \hat{k}) + \hat{k} \log \hat{k}) & \text{if} \;\; \hat{y}^\top \hat{y}/\hat{k} > 1 \\ \hat{y}^\top \hat{y} - 2((p - \hat{k}) \log(p - \hat{k}) + \hat{k} \log \hat{k}) & \text{otherwise}\\  \end{cases}$ \\ \hline
\end{tabular}}
\hfill\                           
\captionof{table}{(Top) Sparsity inducing regularized estimators. $\lambda$ and $\gamma$ denote regularization parameters. In this study, we keep $\gamma$ for SCAD and MCP fixed to 3. (Bottom) Model selection criteria. Here and throughout, $\hat{k}$ refers to the estimated support size, $\hat{y}$ the model predictions of $y$, and $p$ is the total number of features. }
\label{table:algorithms}
\end{table}

A more recent body of work has focused on non-asymptotic analyses of model (\ref{linmodel}) in the framework of compressed sensing rather than regression. Here, the sparsity level of $\beta$ is a priori known, and the sensing matrix X is typically drawn from a random ensemble. In this setting, it is possible to establish sharp transitions in the mean square error distortion of the signal vector as a function of measurement density (i.e., asymptotic n/p ratio) \cite{donoho_message-passing_2009}. Necessary and sufficient conditions on the number of samples needed for high probability recovery of the support of $\beta$ by the Lasso was treated in \cite{wainwright_sharp_2009}. Subsequently, a series of works examined the information theoretic limits on sparse support recovery by forgoing analysis of computationally tractable estimators in favor of establishing the sample complexity of exhaustive evaluation of all $\binom{p}{k}$ possible supports via maximum likelihood decoding \cite{m._j._wainwright_information-theoretic_2009, atia_boolean_2009, s._aeron_information_2010, cem_aksoylar_information-theoretic_2014, j._scarlett_compressed_2013, j._scarlett_limits_2017, c._aksoylar_sparse_2017, k._rahnama_rad_nearly_2011}. This approach provides information theoretic bounds on the selection performance of any inference algorithm, and a measure of the suboptimality of existing algorithms. 

Of particular relevance to this work are \cite{m._j._wainwright_information-theoretic_2009} and \cite{j._scarlett_limits_2017}, whose analyses permit correlated sensing (i.e., design) matrices. Let $\beta_{\text{min}}$ be the minimum non-zero coefficient of $\beta$, $\sigma^2$ be the additive noise variance, and $\Sigma$ be the covariance matrix of the distribution from which columns of $\bm{X}$ are drawn. Denote the set of all subsets of $\{1, 2, ..., p\}$ of size k as $\mathcal{I}_k$. $\mathcal{I}_k$ indexes possible model supports. Given $S, T \in \mathcal{I}_k$ we define the matrix $\Gamma(S, T)$ to be the Schur complement of $\Sigma_{S \cup T, S\cup T}$ with respect to $\Sigma_{TT}$:

$$\Gamma(T, S) = \Sigma_{S \setminus T, S \setminus T} - \Sigma_{S \setminus T, T}(\Sigma_{TT})^{-1} \Sigma_{T, S \setminus T}$$

and let $\rho(\Sigma, k)$ be the smallest eigenvalue this matrix can have for any $T$:

\begin{equation}
\rho(\Sigma, k) = \min_{T \in \mathcal{I}_k \setminus S} \lambda_{\text{min}}(\Gamma(T, S))
\end{equation}

From these quantities, we define $\alpha$:
\begin{equation}
\alpha = \frac{\beta_{\text{min}}^2 \rho(\Sigma, k)}{\sigma^2}
\end{equation}

In Theorem 1 of \cite{m._j._wainwright_information-theoretic_2009}, sufficient conditions on the sample size required for an exhaustive search maximum likelihood decoder to recover the true model support with high probability are given in terms of $p$, $k$, and $\alpha$:

\begin{theorem} Theorem 1 of \cite{m._j._wainwright_information-theoretic_2009}. Define the function $g(c_1, p, k, \alpha)$:

$$ g(c_1, p, k, \alpha) := (c_1 + 2048) \max \left\{\log \binom{p - k}{k}, \log(p - k)/\alpha \right\}$$

If the sample size $n$ satisfies $n > g(c_1, p, k, \alpha)$ for some $c_1 > 0$, then the probability of correct model support recovery exceeds $1 - \exp(-c_1(n - k))$.

\end{theorem}

If $\alpha^{-1} > p \log (p - 2k) + 2k/p$, then $g$, and therefore the sample complexity of support recovery, will be modulated by $\alpha$ for $p$ large enough.  Many of the design matrices considered in our numerical study (see Section 3) satisfy this condition. 

In contrast to compressed sensing, the sparsity level of $\beta$ (i.e., $k$) is typically unknown in applications of regression. Furthermore, sufficient conditions on high probability theory such as Theorem 1 above rely on concentration inequalities, which may formally hold in the non-asymptotic setting, but are rarely tight. As a result, the applicability of these results for practitioners evaluating the robustness of support recovery in finite sample regression is unclear. The main contribution of this work is to address this gap through extensive numerical simulations. We find $\alpha$ to be a useful measure of the difficulty of a particular regression problem, and find selection accuracy performance to be modulated by $\alpha$ even when it does not satisfy the condition stated above.

Previous empirical works have evaluated the effects of collinearity on domain specific regression problems \cite{dormann_collinearity_2013, vatcheva_multicollinearity_2016} and evaluate the efficacy of various information critera for model selection \cite{schoniger_model_2014, brewer_relative_2016, dziak_sensitivity_2020}. Finally, the performance scaling of a series of sparse estimators with sample size is evaluated in \cite{bertsimas_sparse_2020}.

In contrast, we specifically consider the differing effects on selection accuracy of \emph{joint} choices of estimators and model selection criteria. We demonstrate that the choice of model selection criteria significantly modulates the selection performance of estimators, and that there are empirically identifiable transition points in the value of $\alpha$ beyond which the selection performance of all inference procedures degrades.

\section{Methods}

\begin{center}
    \includegraphics[width=0.75\textwidth]{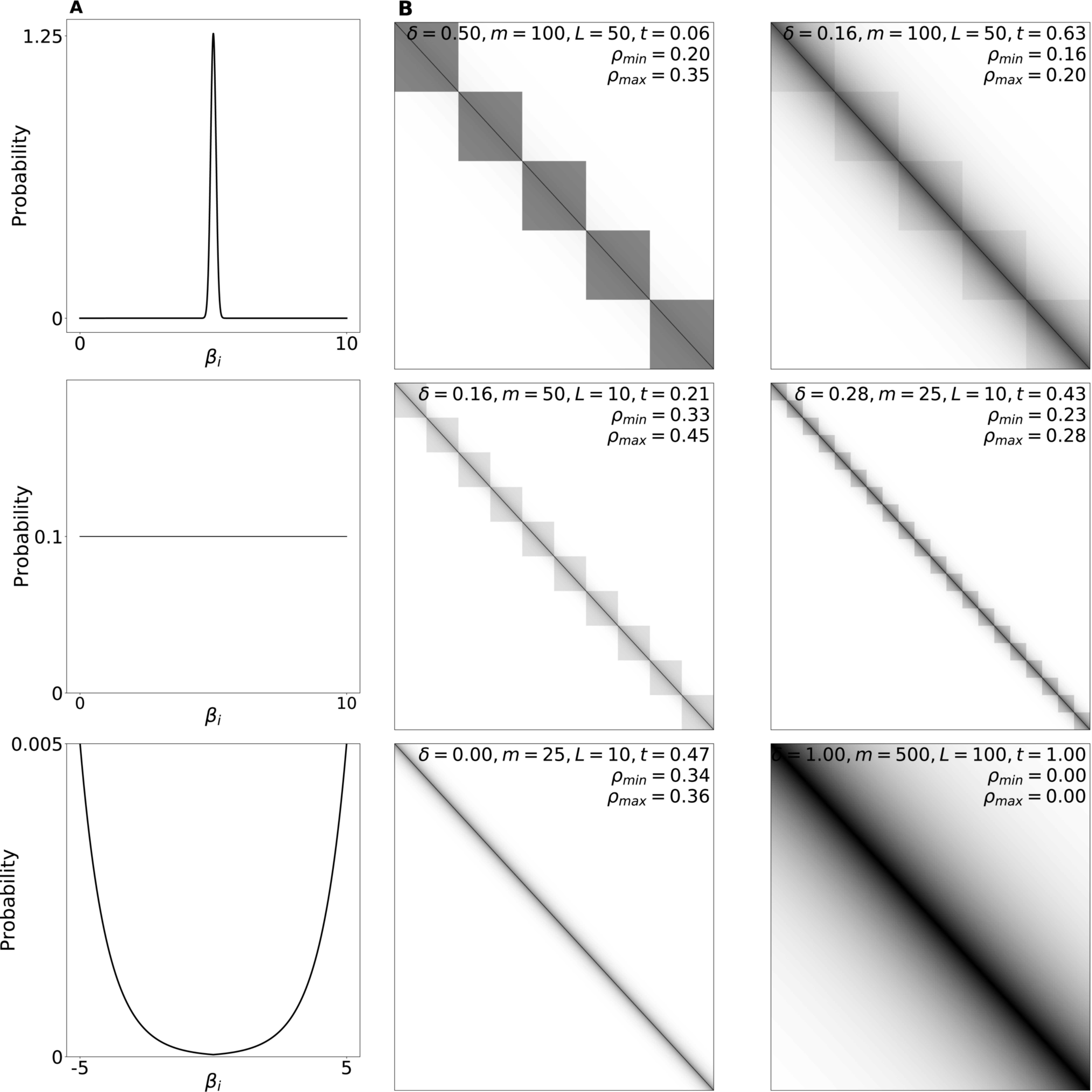}
    \captionof{figure}{Design of Simulation Study. (a) (Right column) Coefficients $\beta$ are drawn from a narrowly peaked Gaussian, uniform, and inverse exponential distribution. (b) (Left column) Design matrices are parameterized as $\Sigma = t \oplus_{i} \delta I_{m \times m} + (1 - t) \Lambda(L)$ where $\Lambda(L)_{ij} = \exp(-|i - j|/L)$ and $I_{m \times m}$ is the m-dimensional identity matrix. Parameters $\delta, m, t$ and $L$ are shown for each example design matrix. Also shown are bounds for the minimum and maximum $\rho(\Sigma, k)$ across $k$.}
    \label{fig:simstudy}
\end{center}

\subsection{Simulation Study}

We consider regression problems with 500 features with 15 different model densities (i.e., $|\beta|_0$) logarithmically distributed from 0.025 to 1. Additionally, we vary over the following design parameters:

\begin{enumerate}
    \item 80 covariance matrices $\Sigma$ of exponentially banded, block diagonal, or a structure that interpolates between the two (see Figure \ref{fig:simstudy}).
    \item Three different $\beta$ distributions: a sharply peaked Gaussian, a uniform, and an inverse exponential distribution (see Figure \ref{fig:simstudy})
    \item Signal to noise (SNR) ratios of 1, 2, 5, 10. We define signal to noise as $|X \beta|_2^2/\sigma^2$.
    \item Sample to feature (n/p) ratios of 2, 4, 8, and 16. 
\end{enumerate}

To simplify the presentation, we often restrict the analysis to the following three combinations of SNR and n/p ratio that represent ideal signal and sample, SNR starved, and sample starved scenarios, respectively:

\begin{enumerate}
    \item Case 1: SNR 10 and n/p ratio 16
    \item Case 2: SNR 1, and n/p ratio 4
    \item Case 3: SNR 5 and n/p ratio 2
\end{enumerate}

A distinct model design is comprised of a particular model density, predictor covariance matrix, a coefficient distribution drawn from one of the three $\beta$-distributions, an SNR, an n/p ratio. Each distinct model is fit over 20 repetitions with each repetition being comprised of a new draw of $X \sim \mathcal{N}(0, \Sigma)$ and $\epsilon \sim \mathcal{N}(0, \sigma^2)$, with $\sigma^2$ set by the desired SNR. We use the term estimator to refer to a particular regularized solution to problem \ref{linmodel} (e.g. Lasso) and model selection criteria to refer to the method used to select regularization strengths (e.g. BIC). The estimators and model selection criteria we consider are listed in Table \ref{table:algorithms}. We use the term inference algorithm to refer to particular choices of estimator and model selection criteria. Fitting with 5 estimators and 5 selection methods, we have run over 28 million fits, requiring over a million computing hours on the National Energy Research Supercomputing Center (NERSC). 

\subsection{Evaluation Criteria}

Let $S = \{i | \beta_i \neq 0\}$ in eq. \ref{linmodel}, and $\hat{S} = \{i | \hat{\beta_i} \neq 0\}$, i.e. the true and estimated model supports. Then, we evaluate regression on the basis of: 

\begin{enumerate}
    \item Selection Accuracy: $1 - \frac{|(S \backslash \hat{S}) \cup (\hat{S} \backslash S)|_0}{|S|_0 + |\hat{S}|_0}$
    \item False Negative Rate: $\frac{|S \backslash \hat{S}|_0}{|S|_0}$
    \item False Positive Rate: $\frac{|\hat{S} \backslash \hat{S}|_0}{p - |S|_0}$
\end{enumerate}

We use $\alpha$ to associate a single scalar to measure the difficulty of a regression problem. Smaller $\alpha$ correspond to harder regression problems. In practice, we do not calculate $\rho(\Sigma, k)$ explicitly, but rather lower bound it (Supplement Section 1). The parameter $\rho(\Sigma, k)$ becomes smaller with larger k.    

\section{Results}
\FloatBarrier
\begin{figure}
    \includegraphics[width=\textwidth]{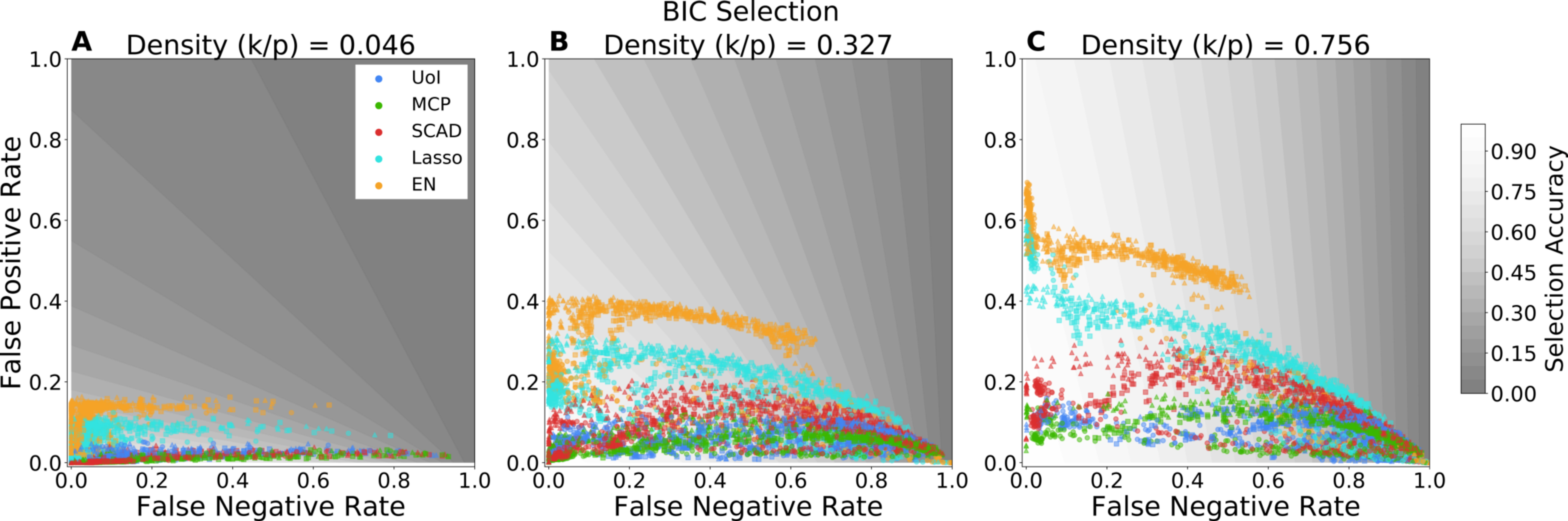}
    \includegraphics[width=\textwidth]{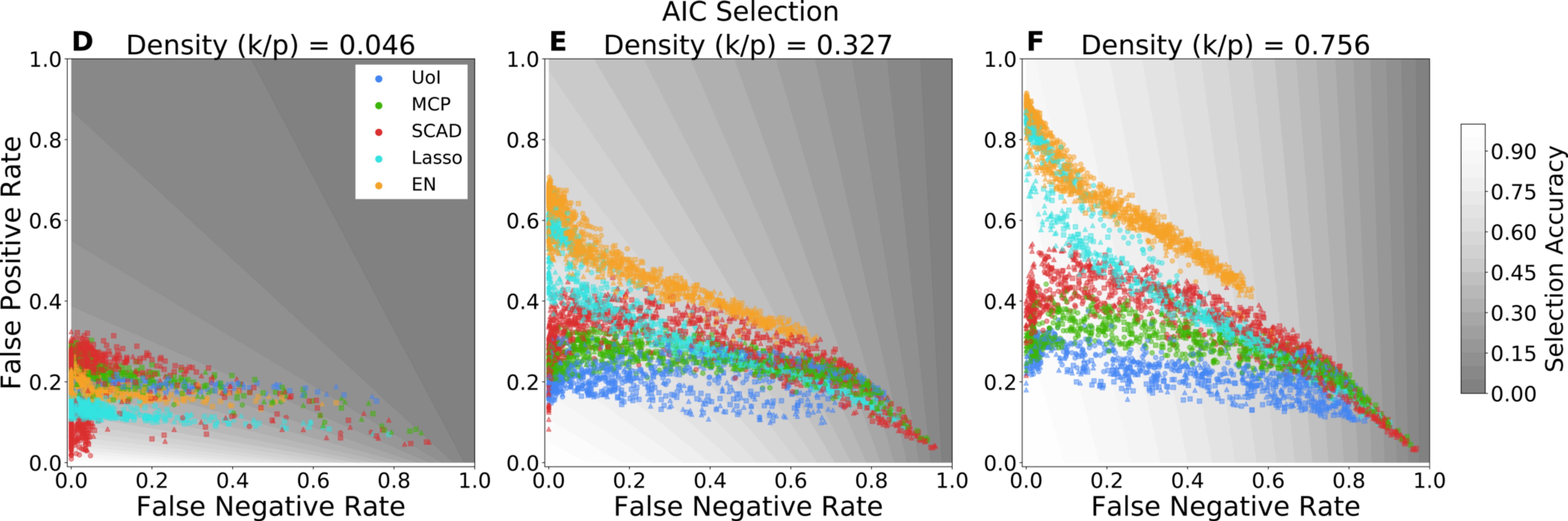}
    \captionof{figure}{Scatter plots of the false negative rate vs. false positive rate for BIC selection (A-C) and AIC selection (D-F) across 3 different model densities (n/p ratio = 4, all signal to noise parameters included). Each scatter point represents a single fit. $\beta$ distributions are encoded in marker shapes (square: uniform distribution, triangular: inverse exponential distribution, circular: Gaussian distribution). Shaded regions represent regions of equal selection accuracy. The orientation of these regions for different model densities illustrates the differing contributions of false negatives vs. false positives, with false positive control being far more important for sparser models, and conversely false negatives being more important for denser models. Estimators can be seen to be characterized by specific tradeoffs between the false positive and false negative control, with SCAD/BIC/UoI (red/green/blue) controlling the false positive rate most aggressively, whereas Elastic Net (orange) controls for false negatives more effectively. The tails of the scatter points extending towards the bottom right of the plot are comprised of the model designs with smallest $\alpha$.}
    \label{fig:scatter}
\end{figure}

\subsection{False Positive/False Negative Characteristics}

We first visualized support selection performance across estimators by scattering the false negative rate vs. false positive rate of each fit for several representative model densities (Figure \ref{fig:scatter} for BIC and AIC selection, Figure \ref{s:fnrfprscatter} for other criteria). Each scatter point represents the selection characteristics of fits to a distinct model design averaged over its 20 instantiations. The boundaries of the grayscale partitions of the false positive false negative rate plane correspond to contours of equal selection accuracy. The rotation of these contours with the true underlying model density reflects the relative importance of false negative and false positive control in modulating selection accuracy. Specifically, rotation towards the horizontal implies larger sensitivity to false positives, while conversely rotation towards the vertical implies greater sensitivity towards false negatives. 

The accuracy of estimators exhibited clear structure that depends on the characteristics of the model design described above. We observe in panel A of Figure \ref{fig:scatter} that estimators that more aggressively promote sparsity (SCAD, MCP, UoI in red, green, and dark blue, respectively) featured better selection accuracy at low model densities (i.e. scatter points for these estimators lie in the white to light gray shaded regions), whereas those that control false negatives less aggressively, namely the Elastic Net (orange) and to a lesser extent the Lasso (cyan), fared better in denser true models (panel C). The scatter points for each estimator formed bands that span the false negative rate. This banding effect was most pronounced for SCAD/MCP/UoI. 

Comparing the BIC selection (Figure \ref{fig:scatter} A-C) to AIC (Figure \ref{fig:scatter} D-F), these scatter plots also revealed that varying model selection methods also systematically shifted false negative/false positive characteristics of estimators. Selection methods with lower complexity penalties (i.e., AIC, CV) lifted the bands up along the false positive direction. Comparing the location of the blue/red/green scatter points between panels B and E, for example, we note that this effect was most dramatic for the set of estimators that most aggressively control false positives (SCAD/MCP/UoI). Consequently, similar tradeoffs as described before arose, with empirically better selection accuracy when models are dense obtained for AIC/CV, and vice versa for larger complexity penalties (BIC). The gMDL and eB methods behaved similarly to BIC (although there are a few exceptions to this, Figures \ref{s:fnrfprscatter}). We conclude that the choice of estimator \emph{and} model selection criteria are both important in determining the false positive/false negative rate behavior of inference strategies.

\subsection{$\alpha$-dependence of False Positives/False Negatives}
\begin{figure}
    \centering
    \includegraphics[width=\textwidth]{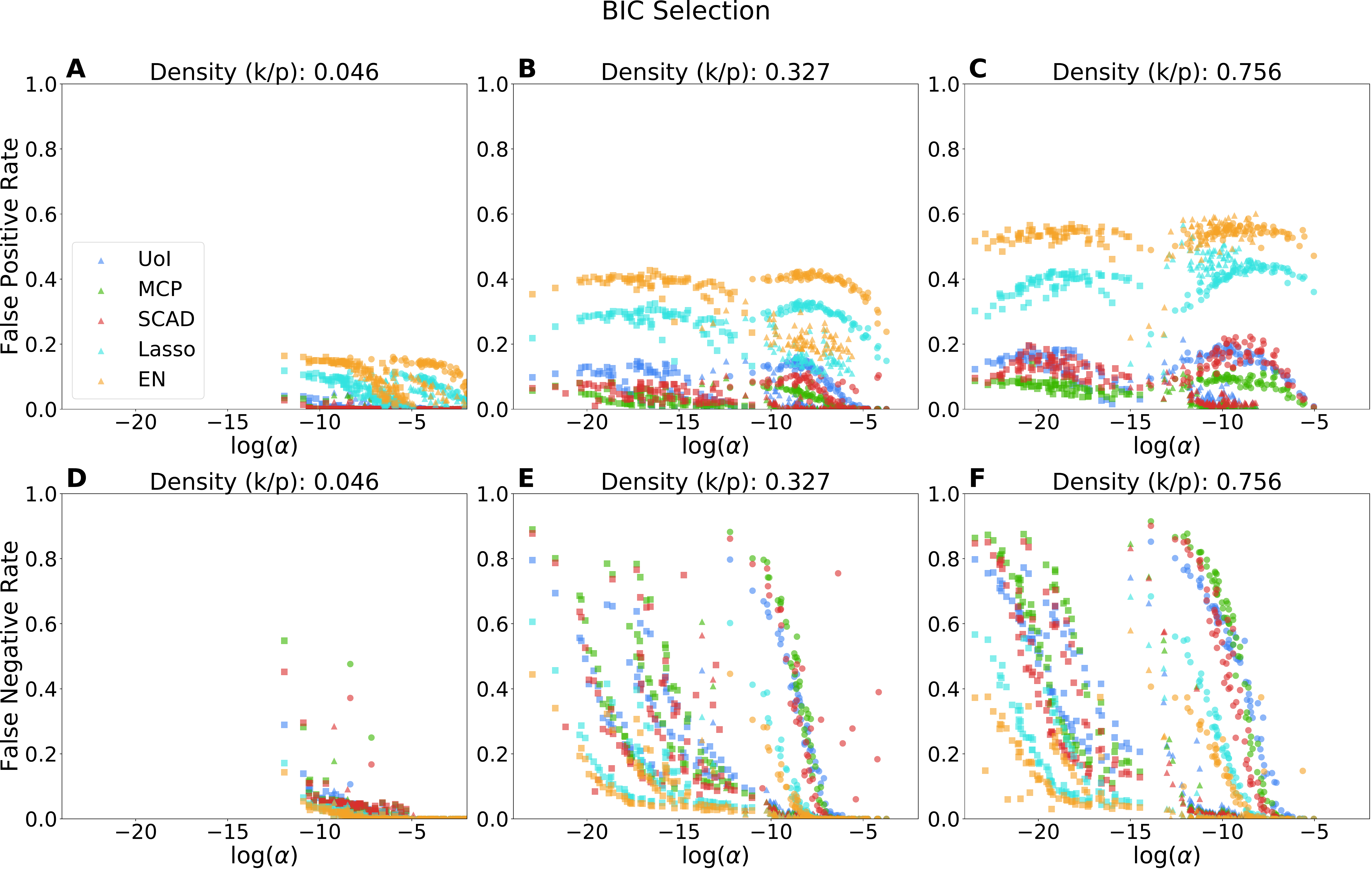}
    \captionof{figure}{Scatter plot of the false positive rate (A-C) and the false negative rate (D-F) vs. $\alpha$ for each estimator using BIC as a selection criteria for three different model densities. $\beta$ distributions are encoded in marker shapes (square: uniform distribution, triangular: inverse exponential distribution, circular: Gaussian distribution). The false positive rate is only weakly modulated by $\alpha$, but is ordered by estimator, with MCP/SCAD/UoI consistently having the lowest false positive rates, Lasso intermediate, and the Elastic Net having the highest false positive rates. The effect becomes pronounced at higher model densities (panel C). For the false negative rates, scatter points follow a characteristic sigmoidal profile, with a region of stable selection accuracy at low correlation (high $\alpha$), followed by an transition point of $\alpha$ after which the selection accuracy montonically decreases. The effect is again most visible at higher model densities. At model density 0.046 (panel D), all estimators are reasonably robust to decreasing alpha.}
    \label{fig:BICalphascatter}
\end{figure}

\begin{figure}
\centering
    \includegraphics[width=\textwidth]{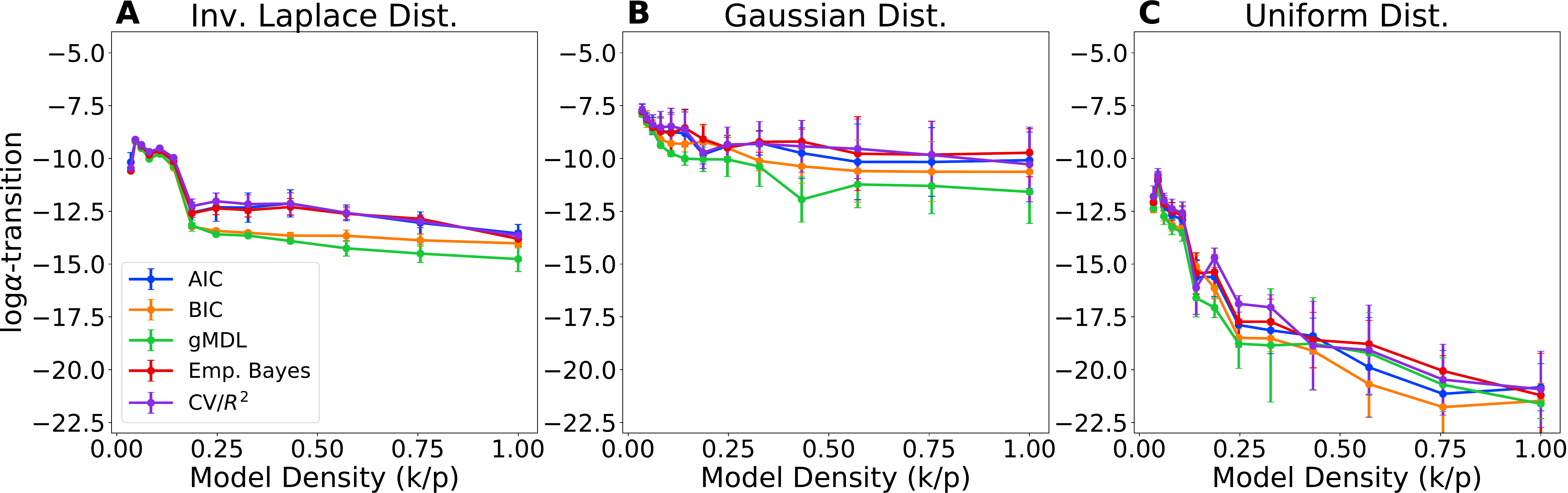}
    \captionof{figure}{Plot of the $\alpha$-transition point associated with an inference algorithm's false negative rate as a function of model density, separated by $\beta$ distribution and selection method. Errorbars are standard deviations taken across repetitions and estimator. The different numerical regimes of the $\alpha$-transition (highest in panel B, intermediate in panel A, and lowest in panel C) is attributable to the different characteristic value of $\beta_{\text{min}}$ for the different $\beta$ distributions.}
    \label{fig:fnralphatransition}
\end{figure}

\begin{figure}
\centering
\includegraphics[scale=0.5]{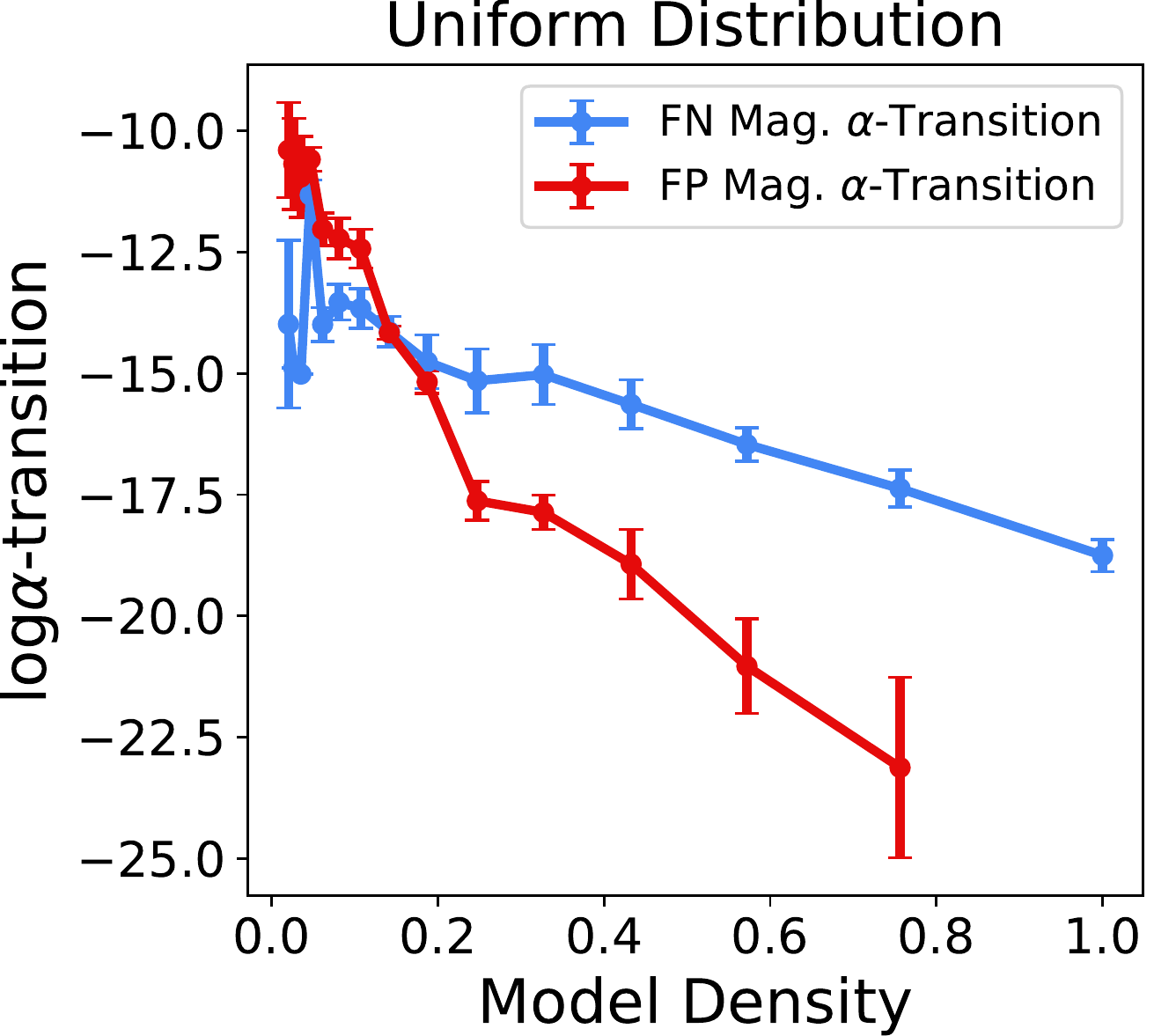}
\captionof{figure}{Plot of the average $\alpha$ transition point for estimation distortion across all inference algorithms and selection methods vs. model density for signal case 1. Errorbars represent standard deviation. After a model density of $>$ 0.15, the transition generally occurs at lower correlations (smaller $\alpha$) for the false negative magnitude. Furthermore, the variance across inference algorithms is consistently smaller for false negatives as opposed to false positives.}
\label{fig:magnitude_transition}
\end{figure}

Recalling that the parameter $\alpha$ tunes the difficulty of the selection problem, we scattered the false positive and false negative rate vs. $\alpha$ for each inference algorithm across different model densities. A representative set of such plots for BIC selection is shown in Figure \ref{fig:BICalphascatter}; other selection methods are shown in Figures S\ref{s:fnr_alphascatter_gMDL}-\ref{s:fnr_alphascatter_CV}. There was broadly large variation in performance modulated by the selection method employed. Furthermore, $\beta$-distributions are separately resolvable due to their different typical values of $\beta_{\text{min}}$. For example, in Figure \ref{fig:BICalphascatter}F, for each estimator, the uniform distribution scatter points (squares) lie to the left of the inverse exponential distribution (triangular), which in turn lies to the left of the Gaussian distribution (circular).  

In line with Figure \ref{fig:scatter}, the false positive rate was not modulated by $\alpha$ (Figure \ref{fig:BICalphascatter} A-C). In fact, for some estimators, the highest false positive rate was achieved for intermediate $\alpha$, followed by a decline in false positive rate for smaller $\alpha$ (e.g. Lasso in Figure \ref{fig:BICalphascatter}C). The false positive rate is instead a characteristic of each estimator. The SCAD/MCP/UoI class of estimators achieved lower false positives than Lasso, which in turn featured lower false positives than the Elastic Net. Model selection criteria can also be classified into a set that led to low false positive rates (gMDL, empirical Bayes, and BIC) vs. those that lead to high false positive rates (AIC, CV), although the Elastic Net with empirical Bayes selection featured the highest false positive rate of any inference algorithm (Supplementary Figure \ref{s:fnr_alphascatter_eb}, panels A-C).

On the other hand, the false negative rate scatter points, when separated by $\beta$-distribution, featured consistent behavior across inference algorithms. Focusing on BIC selection (Figure \ref{fig:BICalphascatter}), all estimators achieved low false negative rates at the low model densities (Figure \ref{fig:BICalphascatter}D). At intermediate model densities (Figure \ref{fig:BICalphascatter}E), the false negative rate remained low until $\log \alpha$ became sufficiently small, at which point it rapidly increases. This value of $\log \alpha$ varied by $\beta$-distribution due to the differing characteristic values of $\beta_{\text{min}}$, occuring around $\log\alpha \approx -7.5$ for the Gaussian distribution at model density 0.327, $\approx \log \alpha = -10$ for the inverse exponential distribution, and $\approx \log \alpha = -15$ for the uniform distribution. Otherwise, this transition point is fairly universal  across inference algorithms.

To produce summary statistics of false negative rates across model densities, selection methods, and n/p ratio/SNR cases, we fit sigmoidal  curves to data for each inference algorithm and for each $\beta$ distribution. The sigmoid curve is described by 4 parameters:

\begin{align*}
    S(\alpha) = c + \frac{a}{1 + \exp(-b (\alpha - \alpha_0))}
\end{align*}

In particular, we use the fitted value for the sigmoid midpoint $\alpha_0$, which we refer to as the $\alpha$-transition point, to quantify the value of $\alpha$ at which false negative rate has begun to increase appreciably. We found a large degree of universality in this transition point across estimators and selection methods. In Figure \ref{fig:fnralphatransition} we have averaged curves across estimators and plotted the mean and standard deviation of the resulting $\alpha$ transition points.  Colors now represent each selection method. The curves for each selection method were strikingly similar within a $\beta$ distribution, with small standard deviations within each selection method indicating universality across estimators. The decrease of the $\alpha$-transition point with increasing model density can be explained by the overall shift of $\alpha$ towards smaller values due to the increase of $\rho(\Sigma, k)$ with $k$.

In the preceding analysis we treated false positives and false negatives as hard thresholded quantities. On the other hand, one can ask whether false negatives primarily arise from setting support elements with small signal strength to zero, and conversely whether false positives are associated with small coefficient estimates. Thus, while exact model support recovery in most cases is unattainable, one would hope that support inconsistencies produce low distortion of the desired coefficient vector. To this end, we calculate the average magnitude of false negatives and false positives, and normalize these quantities by the average magnitude of ground truth $\beta$. Raw scatter plots of these quantities ordered by $\alpha$ can be found in Figures \ref{s:BIC_fnfp}-\ref{s:cv_fnfp}. In the case of false negative magnitudes, we focus on the uniform $\beta$ distribution, as this provides the most ``edge'' cases of small coefficient magnitudes. We found that at low correlations, the hoped for low distortion effect largely holds true, but that there is an $\alpha$ transition point for both false negative and false positives after which significantly larger ground truth $\beta_i$ are selected out, and erroneously selected $\beta_i$ are assigned much larger values relative to the true signal mean. This transition point was again universal across all inference strategies (panels E, F, within each of Figures \ref{s:BIC_fnfp}-\ref{s:cv_fnfp})

In Figure \ref{fig:magnitude_transition}, we plot the transition point as a function of model density averaged across all estimators, selection criteria, and fit repetitions. For model densities $> 0.15$, the transition point occurs at much smaller correlation strengths for the false negative distortions than the false positive distortions. Furthermore, the variance in the location of this transition point for false negative distortions is much smaller than for false positive distortions. For dense, correlated, models in cases 2 and 3 (SNR starved and sample starved, respectively), we find that the mean false positive magnitude can be as high as 5-10 times the true signal mean (e.g. Figure \ref{s:BIC_fnfp}C). 

Overall, these results highlight the usefulness in the parameter $\alpha$, which emerges out of tail bounds on the performance of the exhaustive maximum likelihood decoder, as a quantifier of the difficulty of a sparse regression problem. The value of $\alpha$ at which the false negative rate of inference strategies begins to degrade was found to be universal. A similar universal transition point was found in the value of $\alpha$ at which false negatives and false positives begin to lead to large distortion of in magnitude of non-zero $\beta$ coefficients. 

\subsection{Overall Selection Accuracy}
\begin{figure}
    \includegraphics[width=\textwidth]{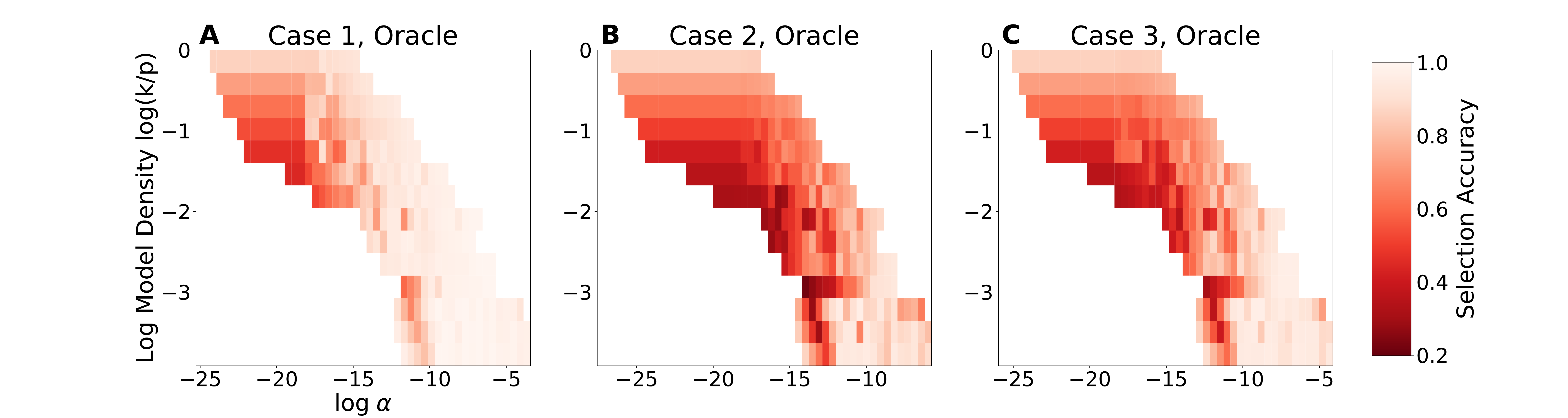}
    \captionof{figure}{Oracle selection accuracy as a function of the log model density and $\alpha$ for each of the 3 signal cases described in Section 3. Each pixel in the colormap is the maximum oracle performance across all estimators for the particular combination of density and $\alpha$. For ideal signal characteristics in Case 1 (panel A), near perfect support recovery is in principle possible for a broad range of correlation strengths for models with density $<$ 0.15. The similar oracle selection accuracies between cases 2 and 3 (panels B and C) suggest that the sample starved and signal starved regression problems behave similarly. As compared to Case 1, worst case performance for intermediate model densities $> 0.1$ and $< 0.5$ is lower, especially for large correlations. For the densest models ($> 0.5$), oracle performance is relatively insensitive to correlation strength, reflecting the insensitivity of the FPR to $\alpha$. Near-perfect support recovery is empirically still possible for the sparsest models (density $< 0.05$).}
    \label{fig:oracle_plot}
\end{figure}

An inference algorithm deployed in practice must employ both an inference estimator and model selection criteria. We have therefore determined what the best performing combination is as a function of underlying model density and $\alpha$. To set an overall scale for these comparisons, one can use an oracle selection criteria that simply chooses the support along a regularization path of maximum selection accuracy. For each value of $\alpha$ and model density, the maximum of this oracular selection across all estimators gives a proxy for the best achievable selection accuracy in principle at finite sample size and SNR. 

In Figure \ref{fig:oracle_plot} we plot the oracle selector for each signal case. In the ideal signal and sample size case (case 1), the oracle selector was able to achieve near perfect selection accuracy in the fully dense models (top row, panel C) and those models with with density $< 0.14$ (log model densities $<-2$) even in model designs with very small $\alpha$. The oracle selector suffered  moderate loss of selection accuracy in intermediate model densities for model designs with small $\alpha$ (darker orange regions of panel C). A similar structure is present in the adequate sample but high noise and low sample size but adequate SNR cases (cases 2 and 3 in panels B and C, respectively), but the magnitude of selection accuracy performance loss and regions of $\alpha$ and model densities for which the loss occurred expanded. In particular, only in the very sparsest models (density $< 0.05$, log model density $< -3$) with larger $\alpha$ was perfect selection possible in principle.

\begin{table}
    \begin{minipage}{0.49\linewidth}
    \sisetup{round-mode=places, round-precision=3}
   \begin{adjustbox}{width=\linewidth, center}
    \begin{filecontents*}{pivoted_table.csv}
sprasity,Case,alg,AIC,BIC,CV,empirical_bayes,gMDL
dense,1,EN,22.595712500288176,15.357086881001472,21.30476942965595,13.346468575187751,11.66836575346861
dense,1,Lasso,20.212996062545937,10.947847081786993,18.151387883250756,8.921314711700681,8.636177884191374
dense,1,MCP,24.45526223742616,5.059093859832466,6.754146924625775,6.695229257159846,11.5459634993642
dense,1,SCAD,22.07039480107169,5.019849111850497,9.135225302647658,5.88364342775043,9.905064981044491
dense,1,UoI Lasso,21.72934864719504,7.764771732319658,31.732772264106753,9.614681590587695,7.831501520925218
dense,2,EN,29.93976056863236,18.539446134291033,25.555838991950157,16.69110011146655,13.502784767316589
dense,2,Lasso,26.463527695721385,14.120111410536826,22.74893354015336,9.593258872945391,8.964677435261677
dense,2,MCP,30.788990478444607,3.8792991332108717,5.971322610959988,4.658561285064508,8.01282729164967
dense,2,SCAD,31.57890465579565,3.4851112186957702,8.154172701563756,4.213284761828551,6.433780574226653
dense,2,UoI Lasso,27.150665372712357,7.286566350522436,36.58773262517934,9.150201752592789,7.023943798201164
dense,3,EN,35.13897508912831,25.14604577316163,33.09804453145588,17.533424758987255,15.37098186357564
dense,3,Lasso,30.622419396093733,18.40193724571007,35.22029596991529,4.600507681980771,5.045515541531781
dense,3,MCP,30.319472917525392,1.1206043216882668,7.51110644486694,1.033189539087383,2.184452604249281
dense,3,SCAD,21.266799460154214,0.814633307087049,9.36148582610094,0.7276493990736207,1.7559286575383344
dense,3,UoI Lasso,29.55763909948862,3.2901718503126207,44.522238993617826,3.07673248043131,3.3956160825214416
full,1,EN,18.289701965738185,26.08672417785288,15.338620038661205,10.420093653260478,12.984576132290886
full,1,Lasso,19.423854451334286,19.653248750276582,17.954545134680156,17.631960110573488,15.227449297484847
full,1,MCP,23.589781922223068,16.525897680797264,14.600089873228937,17.2109873113548,18.155700528929753
full,1,SCAD,21.125068610165012,15.240903995963365,14.118698680847018,15.007219807923367,15.873144261093229
full,1,UoI Lasso,22.029774494140465,19.866396605768628,24.07950045737687,17.420272034274603,15.268411935656092
full,2,EN,22.615104999304457,22.472796534760754,18.381845338791546,13.091869528302421,13.580620850127671
full,2,Lasso,22.495447891329782,19.004132820711085,19.52441324424451,16.7083602040707,14.184892427047892
full,2,MCP,26.410570050502475,13.52050593541502,11.868589422075082,14.38231702752041,14.670516042453833
full,2,SCAD,26.452730927820717,11.789498509559573,12.002631438832378,12.627654186197674,12.265613441787906
full,2,UoI Lasso,23.36638975685683,17.504613919989772,27.575310240416652,15.958870078719553,13.350563750492512
full,3,EN,27.00030017973359,19.228070698399556,25.213785091241846,14.483494996933505,11.964245746452459
full,3,Lasso,23.86722321349259,14.634458430755384,27.151341522954485,5.839744855858705,5.98217887924011
full,3,MCP,23.408383621163402,4.3246115477303535,6.948168483535652,4.716956514368954,5.219505465755397
full,3,SCAD,16.947122469098765,3.2331166877674127,8.051221615328076,3.5341141927404687,4.0386268364385565
full,3,UoI Lasso,22.16261878190285,5.65851226883634,33.79502999411367,5.019856637109265,5.134279617168758
sparse,1,EN,22.595712500288176,15.357086881001472,21.30476942965595,13.346468575187751,11.66836575346861
sparse,1,Lasso,20.212996062545937,10.947847081786993,18.151387883250756,8.921314711700681,8.636177884191374
sparse,1,MCP,24.45526223742616,5.059093859832466,6.754146924625775,6.695229257159846,11.5459634993642
sparse,1,SCAD,22.07039480107169,5.019849111850497,9.135225302647658,5.88364342775043,9.905064981044491
sparse,1,UoI Lasso,21.72934864719504,7.764771732319658,31.732772264106753,9.614681590587695,7.831501520925218
sparse,2,EN,29.93976056863236,18.539446134291033,25.555838991950157,16.69110011146655,13.502784767316589
sparse,2,Lasso,26.463527695721385,14.120111410536826,22.74893354015336,9.593258872945391,8.964677435261677
sparse,2,MCP,30.788990478444607,3.8792991332108717,5.971322610959988,4.658561285064508,8.01282729164967
sparse,2,SCAD,31.57890465579565,3.4851112186957702,8.154172701563756,4.213284761828551,6.433780574226653
sparse,2,UoI Lasso,27.150665372712357,7.286566350522436,36.58773262517934,9.150201752592789,7.023943798201164
sparse,3,EN,35.13897508912831,25.14604577316163,33.09804453145588,17.533424758987255,15.37098186357564
sparse,3,Lasso,30.622419396093733,18.40193724571007,35.22029596991529,4.600507681980771,5.045515541531781
sparse,3,MCP,30.319472917525392,1.1206043216882668,7.51110644486694,1.033189539087383,2.184452604249281
sparse,3,SCAD,21.266799460154214,0.814633307087049,9.36148582610094,0.7276493990736207,1.7559286575383344
sparse,3,UoI Lasso,29.55763909948862,3.2901718503126207,44.522238993617826,3.07673248043131,3.3956160825214416
\end{filecontents*}
\csvreader[tabular=|l|l|l|l|l|l|,
    table head=\hline & \multicolumn{5}{c|}{Selection Method} \\\hline Estimator & AIC & BIC & CV/$R^2$ & Emp. Bayes & gMDL\\\hline,
    filter expr={test{\ifcsvstrcmp{\sparsity}{full}}
                and test{\ifnumequal{\case}{3}}},
    late after last line=\\\hline]
    {pivoted_table.csv}{sprasity=\sparsity, Case=\case, alg=\alg, AIC=\AIC,
                        BIC=\BIC, CV=\CV, empirical_bayes=\empb, gMDL=\gMDL}
                        {\alg & \BoldCaseThreeAll{\AIC} & $\BoldCaseThreeAll{\BIC}$ &
                        \BoldCaseThreeAll{\CV} & \BoldCaseThreeAll{\empb} & \BoldCaseThreeAll{\gMDL}}
    \end{adjustbox}
    \end{minipage} 
    \begin{minipage}{0.49\linewidth}
    \sisetup{round-mode=places, round-precision=3}
   \begin{adjustbox}{width=\linewidth, center}
    \csvreader[tabular=|l|l|l|l|l|l|,
    table head=\hline & \multicolumn{5}{c|}{Selection Method} \\\hline Estimator & AIC & BIC & CV/$R^2$ & Emp. Bayes & gMDL\\\hline,
    filter expr={test{\ifcsvstrcmp{\sparsity}{sparse}}
                and test{\ifnumequal{\case}{3}}},
    late after last line=\\\hline]
    {pivoted_table.csv}{sprasity=\sparsity, Case=\case, alg=\alg, AIC=\AIC,
                        BIC=\BIC, CV=\CV, empirical_bayes=\empb, gMDL=\gMDL}
                        {\alg & \BoldCaseThreeSparse{\AIC} & \BoldCaseThreeSparse{\BIC} & 
                        \BoldCaseThreeSparse{\CV} & \BoldCaseThreeSparse{\empb} & \BoldCaseThreeSparse{\gMDL}}
    \end{adjustbox}
    \end{minipage}
\captionof{table}{Table of summed deviation in selection accuracy from oracular performance. Case 1 Signal Conditions (SNR 10 and n/p 16). (Left) All model densities. (Right) Sparse models only.}
\end{table}

\begin{table}
    \begin{minipage}{0.49\linewidth}
    \sisetup{round-mode=places, round-precision=3}
	\begin{adjustbox}{width=\linewidth, center}
    \csvreader[tabular=|l|l|l|l|l|l|,
    table head=\hline & \multicolumn{5}{c|}{Selection Method} \\\hline Estimator & AIC & BIC & CV/$R^2$ & Emp. Bayes & gMDL\\\hline,
    filter expr={test{\ifcsvstrcmp{\sparsity}{full}}
                and test{\ifnumequal{\case}{2}}},
    late after last line=\\\hline]
    {pivoted_table.csv}{sprasity=\sparsity, Case=\case, alg=\alg, AIC=\AIC,
                        BIC=\BIC, CV=\CV, empirical_bayes=\empb, gMDL=\gMDL}
                        {\alg & \BoldCaseTwoAll{\AIC} & \BoldCaseTwoAll{\BIC} & \BoldCaseTwoAll{\CV}
                        & \BoldCaseTwoAll{\empb} & \BoldCaseTwoAll{\gMDL}}
	\end{adjustbox}
	\end{minipage}
	\begin{minipage}{0.49\linewidth}
    \sisetup{round-mode=places, round-precision=3}
   \begin{adjustbox}{width=\linewidth, center}
    \csvreader[tabular=|l|l|l|l|l|l|,
    table head=\hline & \multicolumn{5}{c|}{Selection Method} \\\hline Estimator & AIC & BIC & CV/$R^2$ & Emp. Bayes & gMDL\\\hline,
    filter expr={test{\ifcsvstrcmp{\sparsity}{sparse}}
                and test{\ifnumequal{\case}{2}}},
    late after last line=\\\hline]
    {pivoted_table.csv}{sprasity=\sparsity, Case=\case, alg=\alg, AIC=\AIC,
                        BIC=\BIC, CV=\CV, empirical_bayes=\empb, gMDL=\gMDL}
                        {\alg & \BoldCaseTwoSparse{\AIC} & \BoldCaseTwoSparse{\BIC} & 
                        \BoldCaseTwoSparse{\CV} & \BoldCaseTwoSparse{\empb} & \BoldCaseTwoSparse{\gMDL}}
    \end{adjustbox}
	\end{minipage}
\captionof{table}{Table of summed deviation in selection accuracy from oracular performance. Case 2 Signal Conditions (SNR 5 and n/p ratio 2). (Left) All model densities. (Right) Sparse models only.} 
\end{table}

\begin{table}
    \begin{minipage}{0.49\linewidth}
    \sisetup{round-mode=places, round-precision=3}
	\begin{adjustbox}{width=\linewidth, center}
    \csvreader[tabular=|l|l|l|l|l|l|,
    table head=\hline & \multicolumn{5}{c|}{Selection Method} \\\hline Estimator & AIC & BIC & CV/$R^2$ & Emp. Bayes & gMDL\\\hline,
    filter expr={test{\ifcsvstrcmp{\sparsity}{full}}
                and test{\ifnumequal{\case}{1}}},
    late after last line=\\\hline]
    {pivoted_table.csv}{sprasity=\sparsity, Case=\case, alg=\alg, AIC=\AIC,
                        BIC=\BIC, CV=\CV, empirical_bayes=\empb, gMDL=\gMDL}
                        {\alg & \BoldCaseOneAll{\AIC} & \BoldCaseOneAll{\BIC} & 
                        \BoldCaseOneAll{\CV} & \BoldCaseOneAll{\empb} & \BoldCaseOneAll{\gMDL}}
	\end{adjustbox}
\end{minipage}
    \begin{minipage}{0.49\linewidth}
    \sisetup{round-mode=places, round-precision=3}
  \begin{adjustbox}{width=\linewidth, center}
    \csvreader[tabular=|l|l|l|l|l|l|,
    table head=\hline & \multicolumn{5}{c|}{Selection Method} \\\hline Estimator & AIC & BIC & CV/$R^2$ & Emp. Bayes & gMDL\\\hline,
    filter expr={test{\ifcsvstrcmp{\sparsity}{sparse}}
                and test{\ifnumequal{\case}{1}}},
    late after last line=\\\hline]
    {pivoted_table.csv}{sprasity=\sparsity, Case=\case, alg=\alg, AIC=\AIC,
                        BIC=\BIC, CV=\CV, empirical_bayes=\empb, gMDL=\gMDL}
                        {\alg & \BoldCaseOneSparse{\AIC} & \BoldCaseOneSparse{\BIC} & 
                        \BoldCaseOneSparse{\CV} & \BoldCaseOneSparse{\empb} & \BoldCaseOneSparse{\gMDL}}
	\end{adjustbox}
\end{minipage}
\captionof{table}{Table of summed deviation in selection accuracy from oracular performance. Case 3 Signal Conditions (SNR 1 and n/p ratio 4). (Left) All model densities. (Right) Sparse models only.} 
\end{table}

For each estimator and selection criteria combination, we take the sum of deviations of its selection accuracy from the oracular performance shown in Figure \ref{fig:oracle_plot} as a measure of sub-optimality. We divide the analysis into an overall measure of performance over the entire density-$\alpha$ plane, as well as restricting the summation to sparse generative models. The results are summarized in Tables 2-4. Entries are normalized by the number of pixels summed over. The best performing inference algorithms are bolded.

When taken across all model densities, in signal case 1 (Table 2, left), the SCAD with BIC selection and SCAD with empirical Bayesian selection emerged as the best inference algorithms with respect to feature selection. When restricted to low SNR or low sample sizes (cases 2 and 3, tables 2,3, left), these strategies remained amongst the best performing, with the Elastic Net with either empirical Bayesian performing the best in case 2, and cross-validated SCAD/MCP exhibiting robust selection in case 3. When restricting to sparse models only, false positive control becomes paramount, and the Elastic Net was no longer competitive. Instead, the SCAD with BIC or empirical Bayes is near optimal in case 1 (table 2, right), and still the best performing in cases 2 and 3 (table 2, 3, right). MCP exhibited similar performance, with UoI Lasso trailing slightly behind. Thus, in general, the SCAD estimator with BIC or empirical Bayesian model selection led to the most robust algorithm for feature selection.

\section{Discussion}

\subsection{Connections to Prior Work}

Our numerical work corroborates and extends several results from the statistical literature in a non-asymptotic setting. We found the frequently employed cross-validated Lasso to be amongst the worst performing selection strategies. It has been shown that using predictive performance as a criteria for regularization strength selection with the Lasso leads to inconsistent support recovery \cite{leng_note_2006}. A necessary and sufficient condition for asymptotically consistent model selection by the Lasso is for the irrepresentable condition to hold \cite{zhao_model_2006}. In the non-asymptotic setting of this study, we find that the parameter $\alpha$ is a more useful modulator of selection accuracy, and that the irrepresentable constant of \cite{zhao_model_2006} tracks the selection accuracy of Lasso only insofar as it tracks $\alpha$ (Supplement Section 5). We find that the SCAD/MCP and UoI Lasso select model supports more robustly in the presence of correlated design. It is known that the SCAD/MCP do not require any strong conditions on the design matrix for oracular properties to hold \cite{loh_support_2017}, and neither does the BoLasso \cite{bach_bolasso_2008}, upon which the selection logic of UoI is partially based on.

Our work demonstrates that the choice of model selection criteria is as important as the choice of estimator to achieve good selection accuracy. The model selection criteria we have considered can all be categorized as penalized likelihood methods. Cross-validation is known to behave asymptotically like the AIC (\cite{shao_asymptotic_1997}. The magnitude of this complexity penalty can be interpreted as a prior on the model size. We correspondingly find that the BIC performs best in sparse models, whereas the AIC and CV perform best in dense models. The tension between the BIC and AIC has been noted in the literature \cite{yang_can_2005}. The asymptotic selection consistency of using BIC to select SCAD regularization strength has been noted in \cite{wang_tuning_2007}. Our numerical investigations reveal that this remains one of the best extant selection strategies in non asymptotic settings with mild correlated variability as well.

The empirical Bayesian and gMDL procedures were devised with complexity penalties nominally adaptive to the underlying model density. We find that these methods lead to good model selection performance across model densities, but only in ideal signal conditions (i.e. case 1) and low design matrix correlations. There is therefore possible room for methodological development of adaptive complexity penalties. We leave this for future work. 

\subsection{Best Practices in Real Data}

Proper model selection is essential for interpretability of parametric models. While sufficient conditions for model selection are available in the literature, they do not provide actionable results for the practitioner in real data. Our extensive numerical simulations reveal best practices. Non-convex optimization estimators such as the SCAD and MCP generically perform better at selection than the Lasso and Elastic Net when the underlying model is sparse. This in line with both prior numerical work and the understanding that asymptotically, these estimators are oracular selectors \cite{fan_variable_2001}, \cite{zhang_nearly_2010}. Our work reveals that this performance gap remains even as design matrices become increasingly correlated. While the SCAD and MCP are nonconvex problems, recent work has shown that the statistical performance of all stationary points is nearly equivalent \cite{loh_regularized_2015}. Furthermore, development of the optimization algorithms for these estimators has matured to the point where regularization paths for the SCAD and MCP can be computed in the same order of magnitude of time as the Lasso/Elastic Net (see for e.g. \cite{zhao_pathwise_2018}). Our work provides further motivation for the adoption of these algorithms. The $\text{UoI}_{\text{Lasso}}$ algorithm has selection performance competitive with MCP and SCAD in many cases. Furthermore, as we show in Supplemtnal Section 4, the OLS-bagging procedure used in coefficient estimates in UoI leads to lower bias/variance estimates than SCAD/MCP. 
    
There is a tradeoff between false positive and false negative control achieved by model selection strategies. False positive control is largely insensitive to the degree of design correlation. Practitioners seeking tight control of false negatives in model selection may be inclined to use the Elastic Net estimator.  The presence of a number of fairly generic $\alpha$ transition points after which selection accuracy degrades, and false negative/positive magnitude inflates suggests a heuristic criteria that could be estimated from the sample covariance. Specifically, combining empirical estimates of the precision matrix with  empirical estimates of $\beta_{\text{min}}$ and $\sigma^2$ allows one to estimate $\alpha$, and therefore have a rough sense of whether selection and estimation performance is likely to have degraded due to correlated covariates or low signal strength.

\section{Conclusions and Future Work}

Our empirical results reveal that the joint choice of sparse estimator and model selection criteria significantly modulates selection performance. Nevertheless, with the exception of the previously mentioned \cite{wang_tuning_2007}, theoretical results that capture non-asymptotic behavior of regularization strength selection via specific model selection criteria are lacking. 

We found no inference algorithm to be dominant across underlying model density in the presence of correlated covariates, including the nominally adaptive empirical Bayes and gMDL selection criteria. Whether these reflect information theoretic constraints or methodological gaps is a potentially avenue of future work. We also believe our observation of a universal $\alpha$-transition point across false negatives and coefficient distortion to be novel. This phenomena is reminiscent of the well known reconstructability transition in compressed sensing as a function of noise level and sampling density \cite{donoho_message-passing_2009}. An average case analysis of coefficient support distortion as a function of $\alpha$ or other spectral parameters of the design matrix will be the topic of future work.

\section{Acknowledgements}

Funding: AK was supported by an NSF GRFP award. KEB was funded by the Kavli Institute, NIH/NINDS R01NS118648, DOE/ASCR AWD00003162.




\Urlmuskip=0mu plus 1mu\relax
\bibliographystyle{elsarticle-num} 
\bibliography{NumericalSelection}       
\beginsupplement
\section{Bounding $\rho(\Sigma, k)$}

Let $\mathcal{I}_k := \{T | T \subseteq \{1, 2, .., p\}, |T| = k \}$ be the set of all subsets of $\{1, 2, ..., p\}$ of size k. Given sets $S, T \in \mathcal{I}_k$ that index 2 possible model supports of size k, we define the matrix $\Gamma(S, T)$ to be the Schur complement $\Sigma_{S \cup T, S\cup T}$ with respect to $\Sigma_{TT}$:

$$\Gamma(T, S) = \Sigma_{S \setminus T, S \setminus T} - \Sigma_{S \setminus T, T}(\Sigma_{TT})^{-1} \Sigma_{T, S \setminus T}$$
 
Given a true support $S$, the quantity $\rho(\Sigma, k)$ is given by the solution of a discrete optimization problem:

$$\rho(\Sigma, k) = \min_{T \in \mathcal{I}_k \setminus S} \lambda_{\text{min}}(\Gamma(T, S))$$

We derive an easy to calculate approximation to this quantity. First, observing that $\Gamma(S, T)$ is just the inverse of the subblock of the precision matrix $\Sigma^{-1}_{S \setminus T, S \setminus T}$, we seek to bound the largest eigenvalue of this subblock:

$$(\rho(\Sigma, k))^{-1} \leq \max_{T \in \mathcal{I}_k \setminus S} \lambda_{\text{max}}(\Sigma^{-1}_{S \setminus T, S \setminus T})$$

We do this via Brauer-Cassini sets \cite{varga_gersgorin_type_2004}:

\begin{prop}
For an arbitrary $n\times n$ complex matrix $A$ with entries $a_{ij}$, let $R_i = \sum_{j \neq i} |a_{ij}|$. Then, define the Brauer sets $K_i$:
$$K_{ij} = \{z \in \mathbb{C}: |z - a_{ii}| |z - a_{jj}| \leq R_i R_j, i \neq j \} $$
The eigenvalues of $A$ lie within $\bigcup_i K_i$
\end{prop}

To bound specifically the largest eigenvalue of $\Sigma^{-1}_{S \setminus T, S \setminus T}$, we use the following proposition:

\begin{prop}
Let $A \in \mathbb{R}^{n \times n}$ be a positive semidefinite matrix and let $\tilde{A}$ be the the matrix that results from sorting the rows of $|A|_{ij} = |a_{ij}|$ in descending order. Define the truncated row sums $\tilde{R}_i = \sum_{j = 1}^m |\tilde{a}_{ij}|$ where $\tilde{a}_{ij}$ are the entries of $\tilde{A}$. Let $B \in \mathbb{R}^{m \times m}$ be a principal submatrix of $A$. The largest eigenvalue of $B$ is bounded from above by:

$$ \max_{i, j: i \neq j} \left[\sqrt{\tilde{R}_i \tilde{R}_j + \frac{1}{4}(|\tilde{a}_{i0}| - 
|\tilde{a}_{j0}|)^2} + \frac{1}{2}(|\tilde{a}_{i0}| + |\tilde{a}_{j0}|)\right]$$
\end{prop}
\emph{Proof:} Since A is positive semidefinite, by Proposition 1, it follows that the largest eigenvalue of A can be no larger than the rightmost boundary of the rightmost Brauer set on the real axis. As a principal submatrix of a positive semidefinite matrix is also positive semidefinite, this holds analogously for the matrix B and the Brauer sets $\hat{K}_{ij} = \{z \in \mathbb{C}: |z - b_{ii}| |z - b_{jj}| \leq \hat{R}_i \hat{R}_j, i \neq j \}$ where $\hat{R}_i = \sum_{j=1, j \neq i}^m |b_{ij}|$. In Cartesian coordinates, the Brauer set is defined on the real axis by $(x - b_{ii})(x - b_{jj}) = \hat{R}_i \hat{R}_j$. The rightmost root of this equation is given by $\frac{1}{2}(b_{ii} + b_{jj}) + \sqrt{\hat{R_i}\hat{R_j} + \frac{1}{4}(b_{ii} - b_{jj})^2}$ 

By sorting $A$ to obtain $\tilde{A}$, we necessarily have 
\begin{align*}\max_{i, j \in \{1, ..., n \}, i \neq j}\frac{1}{2}(|\tilde{a}_{i0}| + |\tilde{a}_{j0}|) &+ \sqrt{\tilde{R_i}\tilde{R_j} + \frac{1}{4}(|\tilde{a}_{i0}| - |\tilde{a}_{j0}|)^2} \geq \\
&\max_{i, j \in \{1, ..., m\}, i \neq j}\frac{1}{2}(b_{ii} + b_{jj}) + \sqrt{\hat{R_i}\hat{R_j} + \frac{1}{4}(b_{ii} - b_{jj})^2}\end{align*} \qedsymbol

Proposition 2 enables us to bound the largest eigenvalue of a subblock of a matrix of a given size. Depending on the overlap between sets $T$ and $S$, the dimension of the matrix $\Sigma^{-1}_{S \setminus T, S \setminus T}$ will vary. To bound the extremization over all $T \in \mathcal{I}_k$, we rely on the Cauchy interlacing theorem:

\begin{prop}
(Cauchy interlacing theorem) Let $A \in \mathbb{R}^{n \times n}$ be a symmetric matrix and $B \in \mathbb{R}^{m \times m}$ be a matrix obtained from A from an orthogonal projection $P$ onto a subspace of dimension m: $B = P^* A P$. Then, if the eigenvalues of $A$ are ordered as $\lambda_1^A \leq \lambda_2^A \leq ... \leq \lambda_n^A$ and the eigenvalues of $B$ ordered as $\lambda_1^B \leq \lambda_2^B \leq ... \leq \lambda_m^B$, the following inequality holds for all $j \leq m$:

$$ \lambda_j^A \leq \lambda_j^B \leq \lambda^A_{n-m+j}$$
\end{prop}

Since, trivially, $\lambda^A_{n - m + j} \leq \lambda^A_n$, we can bound for the largest eigenvalue of a proper submatrix of dimension $k'$  with the a bound for the largest eigenvalue of subbmatrices of dimension $k > k'$. Therefore, we use the results of Proposition 2 to bound the largest eigenvalue of subblocks of $\Sigma^{-1}$ of dimension k, corresponding to searching over $T$ that are completely disjoint from $S$. Inverting this bound then gives a lower bound on $\rho(\Sigma, k)$.

\section{FNR vs. FPR scatter plots across selection methods}

We include the counterparts of Figure \ref{fig:scatter} for the gMDL, empirical Bayes, and cross-validation selection methods in Figure \ref{s:fnrfprscatter}. We note the qualitative similarity of the profile of scatter points for gMDL selection panels A-C to that of BIC (Figure \ref{fig:scatter}, panels A-C). The gMDL selection method, while nominally sensitive to the underlying model sparsity, gave rise to tight false positive control for all estimators, save for the Elastic Net (orange scatters). In contrast to the BIC at dense model density (panel C, both figures), the gMDL selection criteria provided tighter false positive control for the Lasso (cyan scatter points), at the expense of increased false negatives.

In panels A, B, and D, E of figure \ref{s:fnrfprscatter}, we observe that the gMDL and empirical Bayes selection method led to similar selection profiles for UoI, SCAD, MCP, and Lasso, with nearly all scatter points staying at false positive rates $<$ 0.25. However, we also observe that supports selected by using the Elastic Net, in particular (orange), and other estimators for particular sets of parameters, became very dense (false positive rate $\to 1$) at model density 0.33 and especially model density 0.76 (panel F). This led to overall better selection accuracy (white regions) in denser models. 

In panels G-I, we plot similar scatter points for the cross-validation model selection criteria. We observe a high false positive rate for UoI (blue) that was nearly insensitive to the underlying model density. We therefore recommend that cross-validation is not used as a selection criteria for UoI. Otherwise, we observe a selection profile that that is very similar to that of the AIC (Figure \ref{fig:scatter} D-F), with elevated false positive rates that led to low selection accuracy for sparse underlying models, and good selection accuracy for dense underlying models.

\begin{center}
    \includegraphics[width=\textwidth]{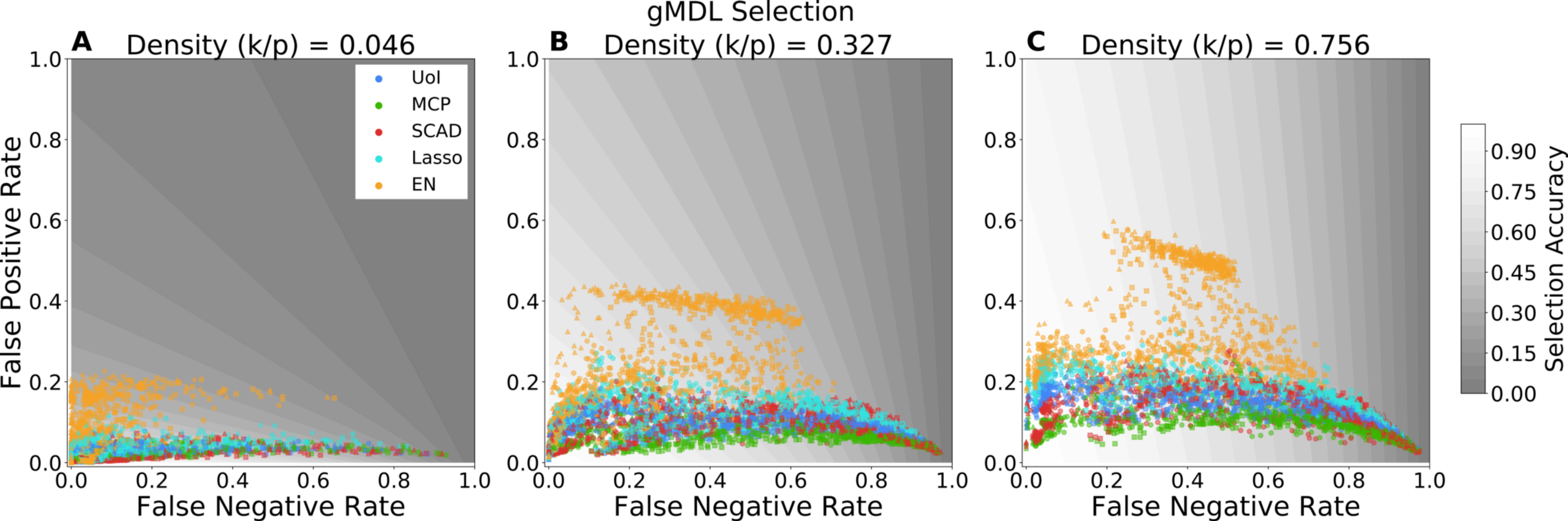}
    \includegraphics[width=\textwidth]{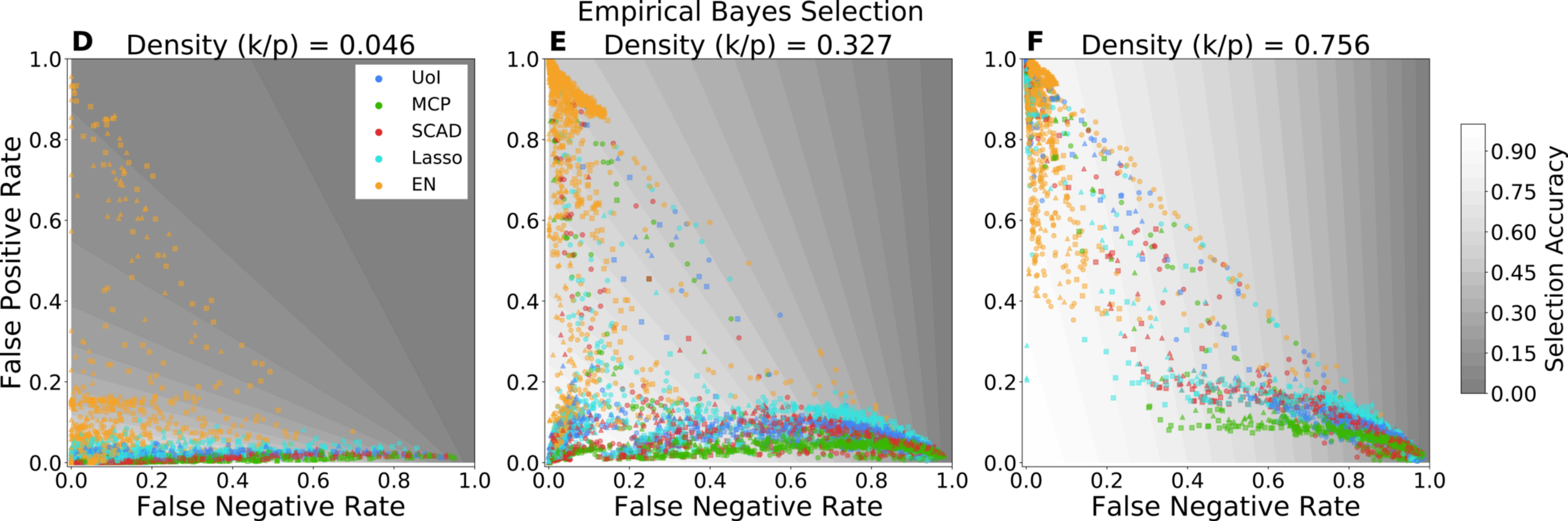}
    \includegraphics[width=\textwidth]{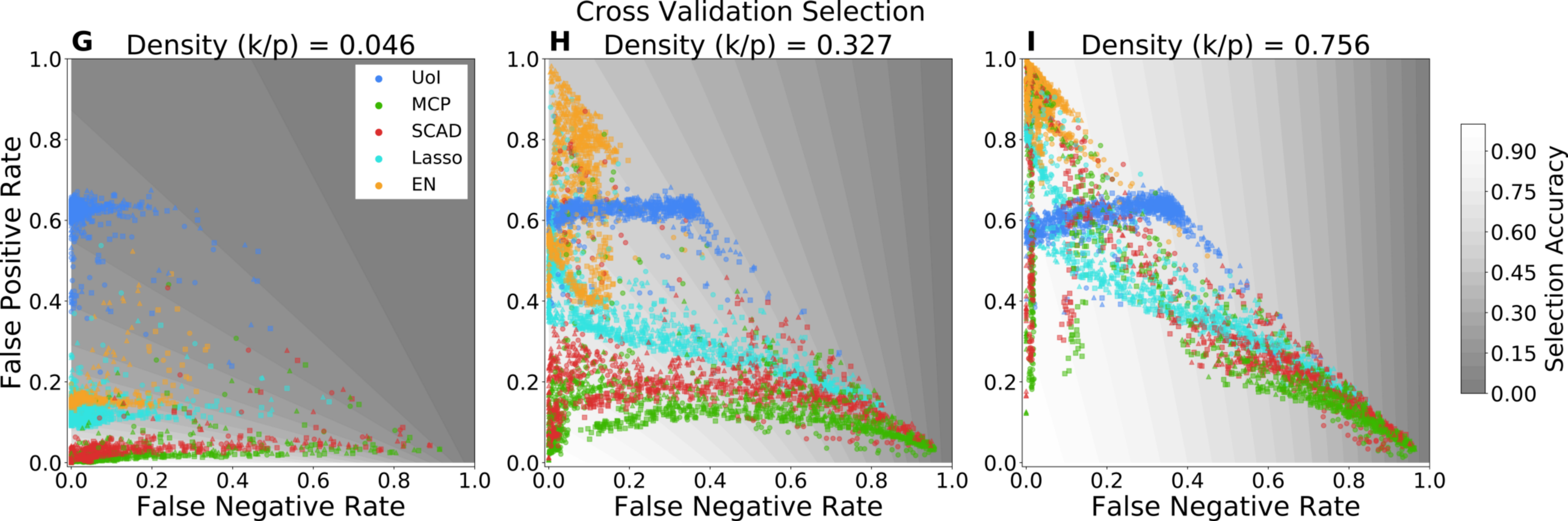}
    \captionof{figure}{Scatter plots of the false negative rate vs. false negative rate for gMDL model selection (panels A-C), empirical Bayes model selection (panels D-F), and cross-validation selection (panels G-I) for 3 different model densities (0.03, 0.33, 0.76). The n/p ratio displayed is 4, all signal to noise parameters are included. Each scatter point represents a single fit. The opacity of the scatter points encodes the signal strength on a normalized scale sensitive to the $\alpha$ value of the fitted regression problem.}
    \label{s:fnrfprscatter}
\end{center}

\section{FPR/FNR vs. alpha scatters}

Here, we include the counterparts of Figure \ref{fig:BICalphascatter} for the other selection methods in Figures \ref{s:fnr_alphascatter_gMDL} - \ref{s:fnr_alphascatter_CV}. As noted in the main text, the false positives did not usually inflate monotonically with decreasing $\log \alpha$. In fact, for the AIC selection method (\ref{s:fnr_alphascatter_AIC} the Elastic Net and Lasso actually exhibited decreasing false positive rates with decreasing $\alpha$ (panels B, C). We also note the very high false positives rates resulting from cross validation selection (Figure \ref{s:fnr_alphascatter_CV}), especially from UoI at low model densities (panels A, B) and Elastic Net at high densities (model C). For model designs with the lowest $\alpha$, the Elastic Net with both cross-validation and empirical Bayes selection tended to include nearly all features. 

A visual inspection of the lower panels (D-F) of the scatter plots below suggests highlights the universal $\alpha$ transition in false negative rates discussed in the main text. Nevertheless, some finer structure is observable. As model densities increases (panel F of all figures), the Elastic Net achieved lower false negative rates and was more robust to small $\alpha$, with its transition point occuring for smaller $\alpha$. Additionally, for empirical Bayes and cross-validation selection (Figures \ref{s:fnr_alphascatter_eb} and \ref{s:fnr_alphascatter_CV}), the Elastic Net had almost no false negatives for all $\alpha$ and model densities. We finally note that the inverse exponential distribution (triangular points) induced very false negatives by any inference algorithms, likely due to its coefficient magnitudes being concentrated towards larger values. 

\begin{center}
    \includegraphics[width=\textwidth]{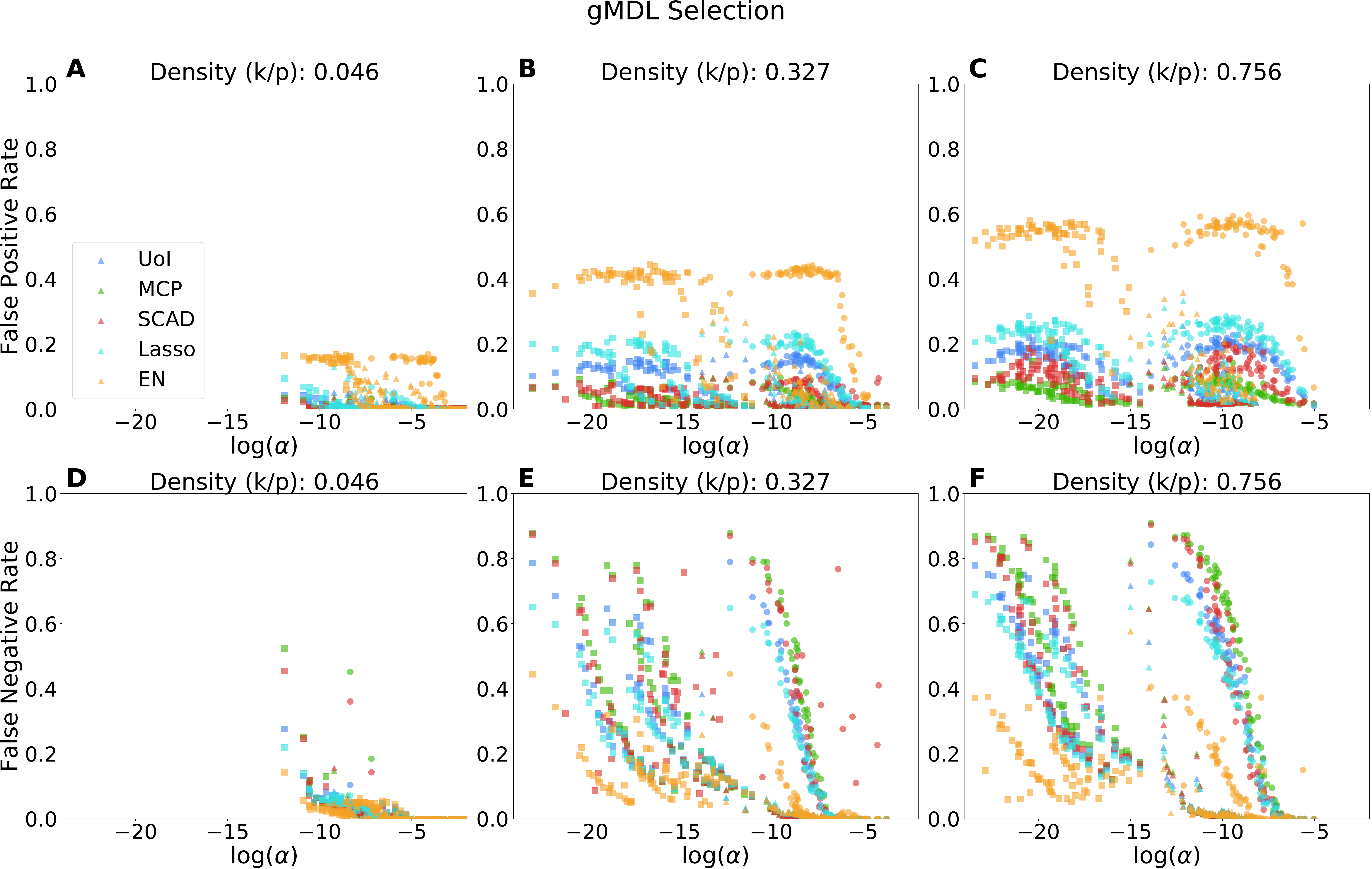}
    \label{s:fnr_alphascatter_gMDL}
    \captionof{figure}{Panels A-C: Plot of the false positive rate vs. $\log \alpha$ for signal case 3 (high FNR, n/p ratio) across several model densities. gMDL selection method was employed.}

    \includegraphics[width=\textwidth]{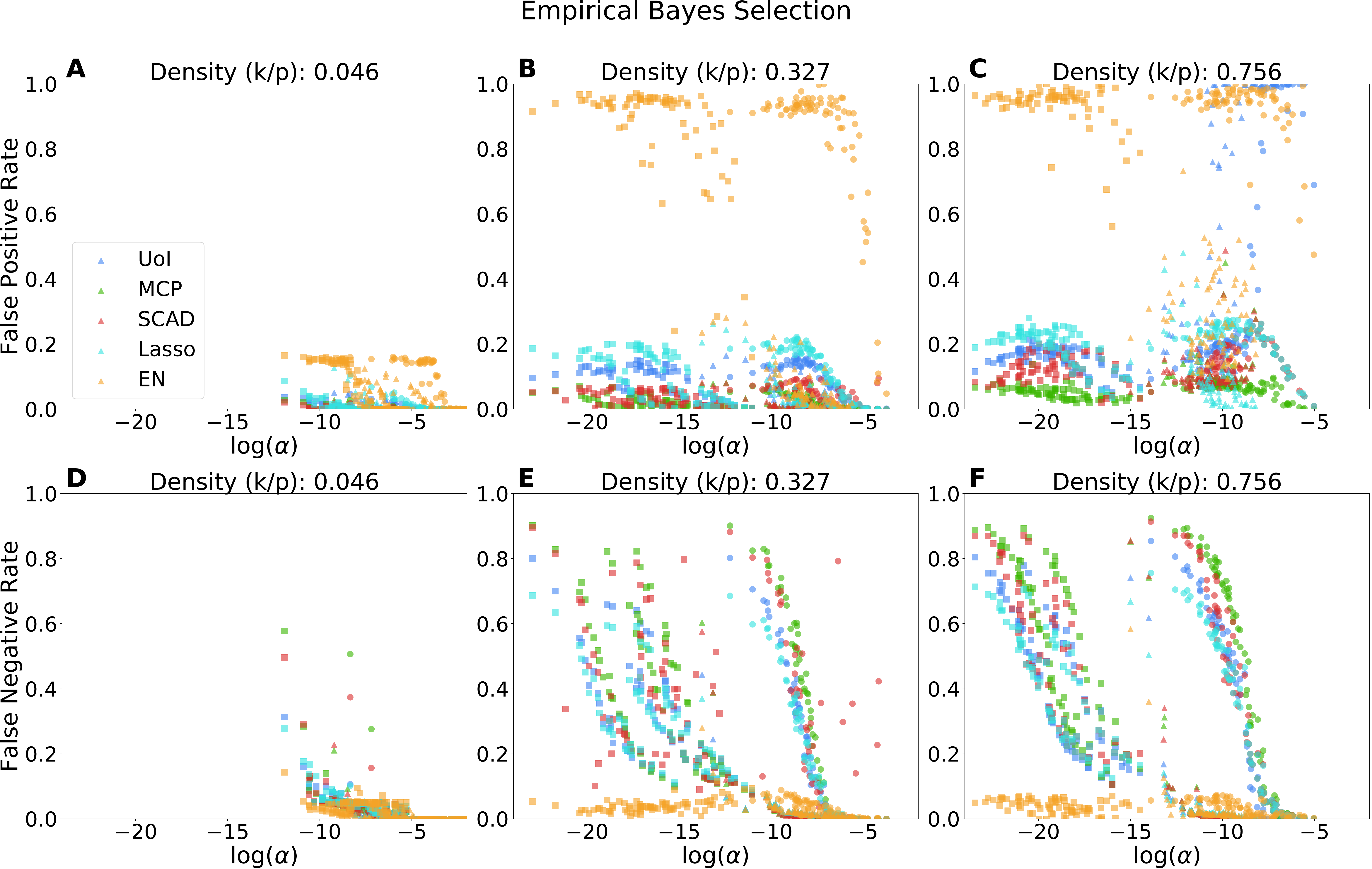}
    \label{s:fnr_alphascatter_eb}
    \captionof{figure}{Panels A-C: Plot of the false positive rate vs. $\log \alpha$ for signal case 3 (high FNR, n/p ratio) across several model densities. Empirical Bayes selection method was employed.}

    \includegraphics[width=\textwidth]{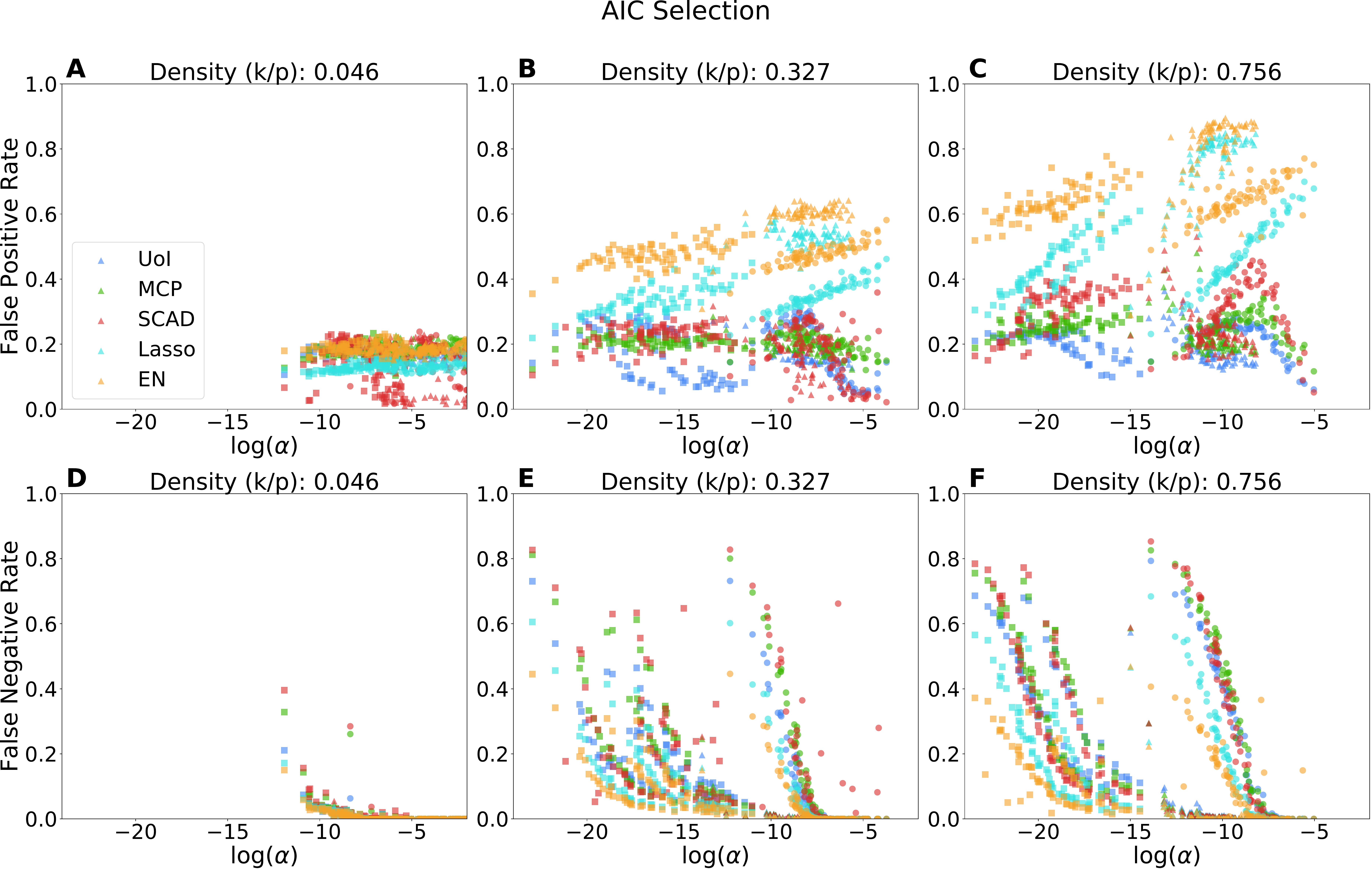}
    \captionof{figure}{Panels A-C: Plot of the false positive rate vs. $\log \alpha$ for signal case 3 (high FNR, n/p ratio) across several model densities. AIC selection method was employed.}
    \label{s:fnr_alphascatter_AIC}

    \includegraphics[width=\textwidth]{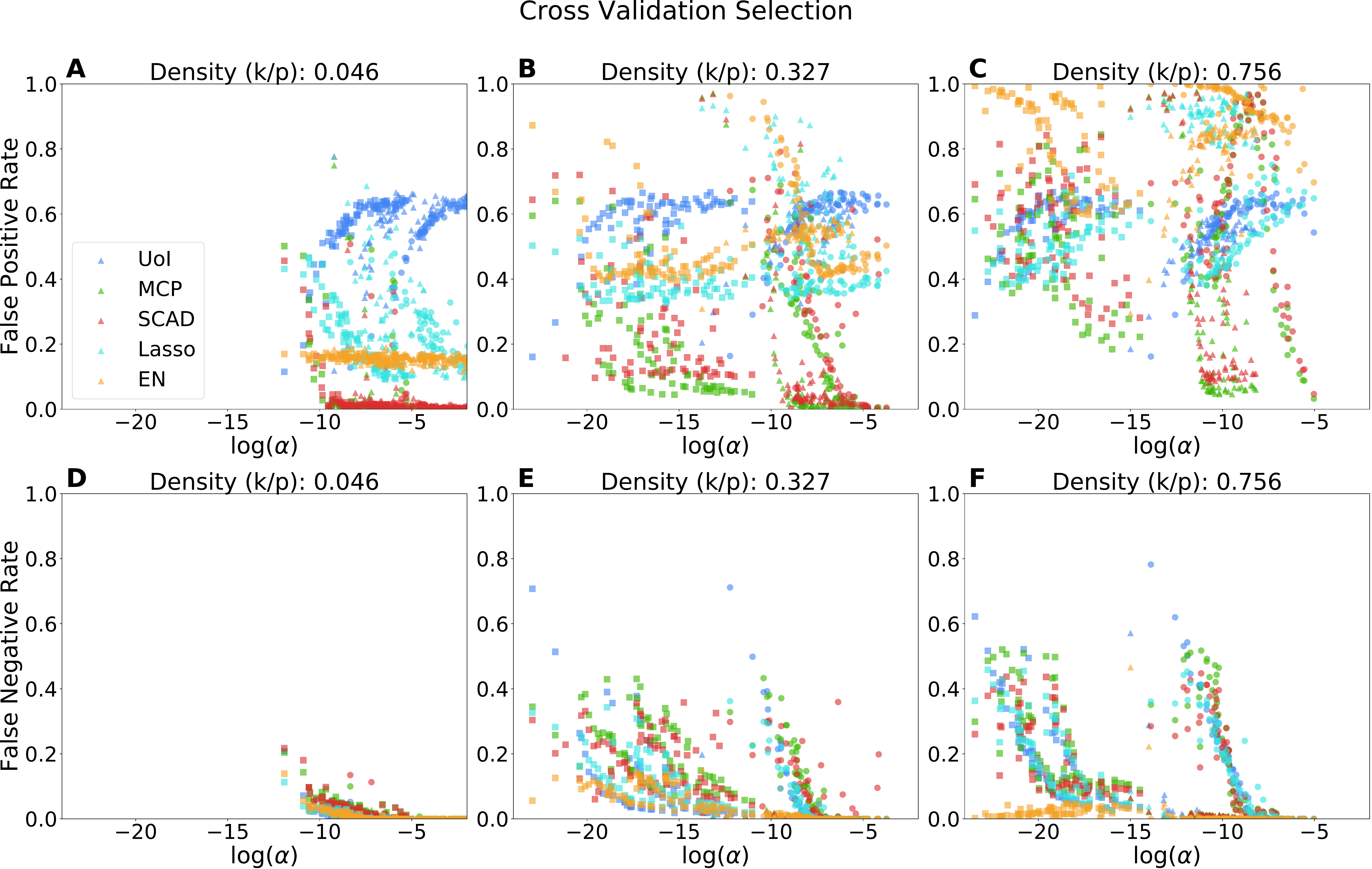}
    \captionof{figure}{Panels A-C: Plot of the false positive rate vs. $\log \alpha$ for signal case 3 (high FNR, n/p ratio) across several model densities. Cross Validation selection method was employed.}
    \label{s:fnr_alphascatter_CV}
\end{center}

\section{FN/FP magnitude scatters}

In order to assess the distortion of the true coefficient vector associated with the magnitude assigned to false positives and the magnitude of false negatives, we scatter in Figures \ref{s:BIC_fnfp}-\ref{s:cv_fnfp} the average magnitude of false positives relative to the mean of the non-zero $\beta$, and similarly the average magnitude of false negatives relative to the mean of the non-zero $\beta$. The latter quantity must remain between 0 and 1. Fits to these scatter points are used to generate Figure \ref{fig:magnitude_transition}. There is largely comparable behavior across selection methods. Taking the BIC as an example, in panels A-C of figure \ref{s:BIC_fnfp}, for the lowest model density, average false positive magnitude remained small for a wide range of $\alpha$, with a nascent transition to magnitudes comparable to the signal mean present around $\log \alpha \approx -12.5$. As the model density increases (panel B), a clear transition point similar across all algorithms was present after which false positives are estimated to have increasingly larger values. It is notable, looking at panel C across the figures, that the SCAD and MCP had the largest values assigned to false positives. 

For the average magnitude of false negatives, again taking BIC as an example (panels D-F of Figure \ref{s:BIC_fnfp}), we again see universality across algorithms in the value of $\alpha$ at which the increasingly larger coefficient values were erroneously estimated to be zero. Comparing panel F to D, we see that this effect is more pronounced in denser models. Overall then, the distortion of the estimated $\beta$ vector via the model estimation strategies considered here became more severe as the underlying model density increases.

\begin{center}
    \includegraphics[width=\textwidth]{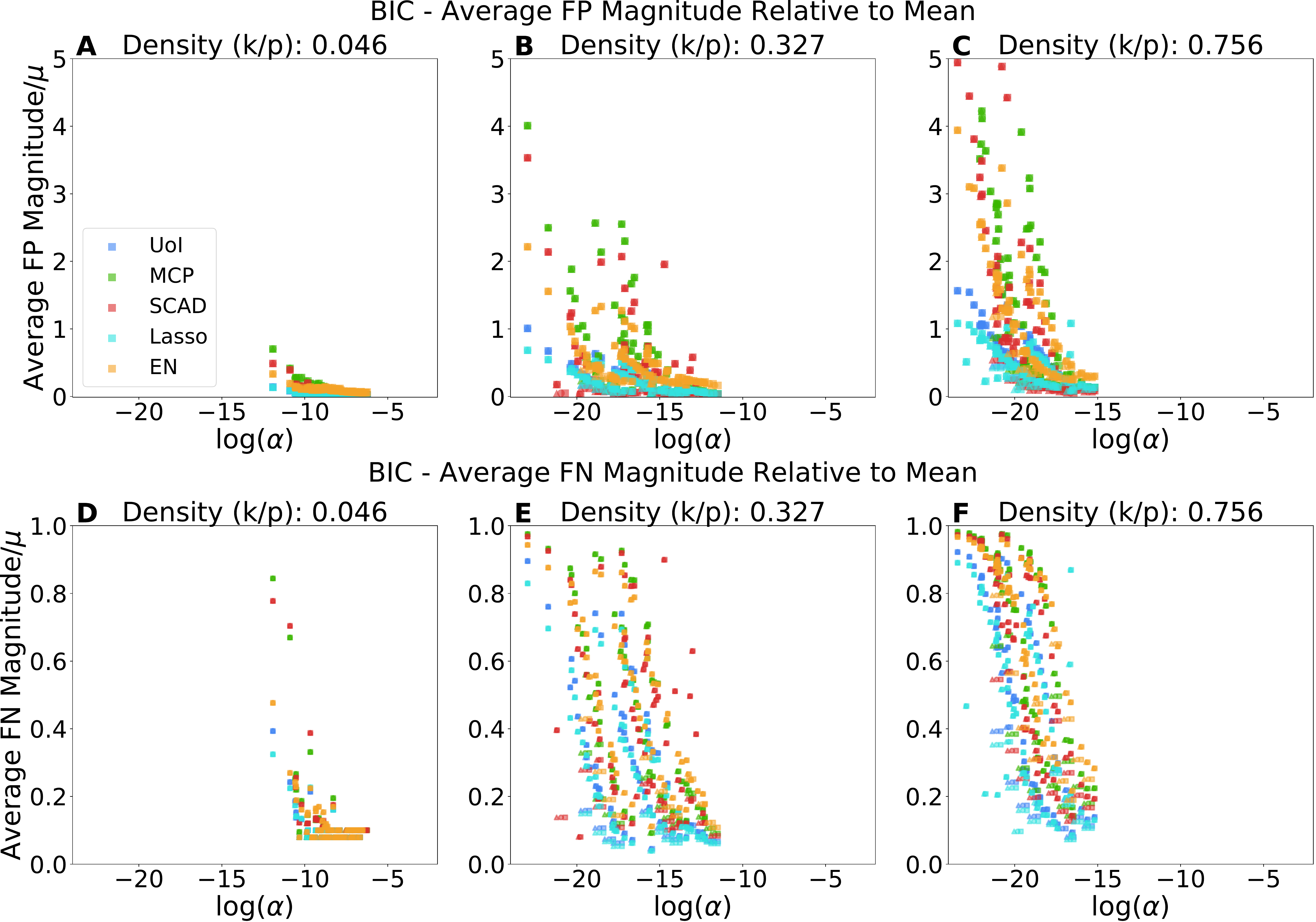}
    \captionof{figure}{Panels A-C: Average magnitude of false positives divided by the mean of the non-zero $\beta$ coefficients for gMDL selection. Panels D-F: Average magnitude of false negatives divided by the mean of the non-zero $\beta$ coefficients. Results for only the uniform distribution $\beta$ distribution shown. For this and subsequent figures, the n/p ratio is 16 and the signal to noise ratio is 10.}
    \label{s:BIC_fnfp}
    \includegraphics[width=\textwidth]{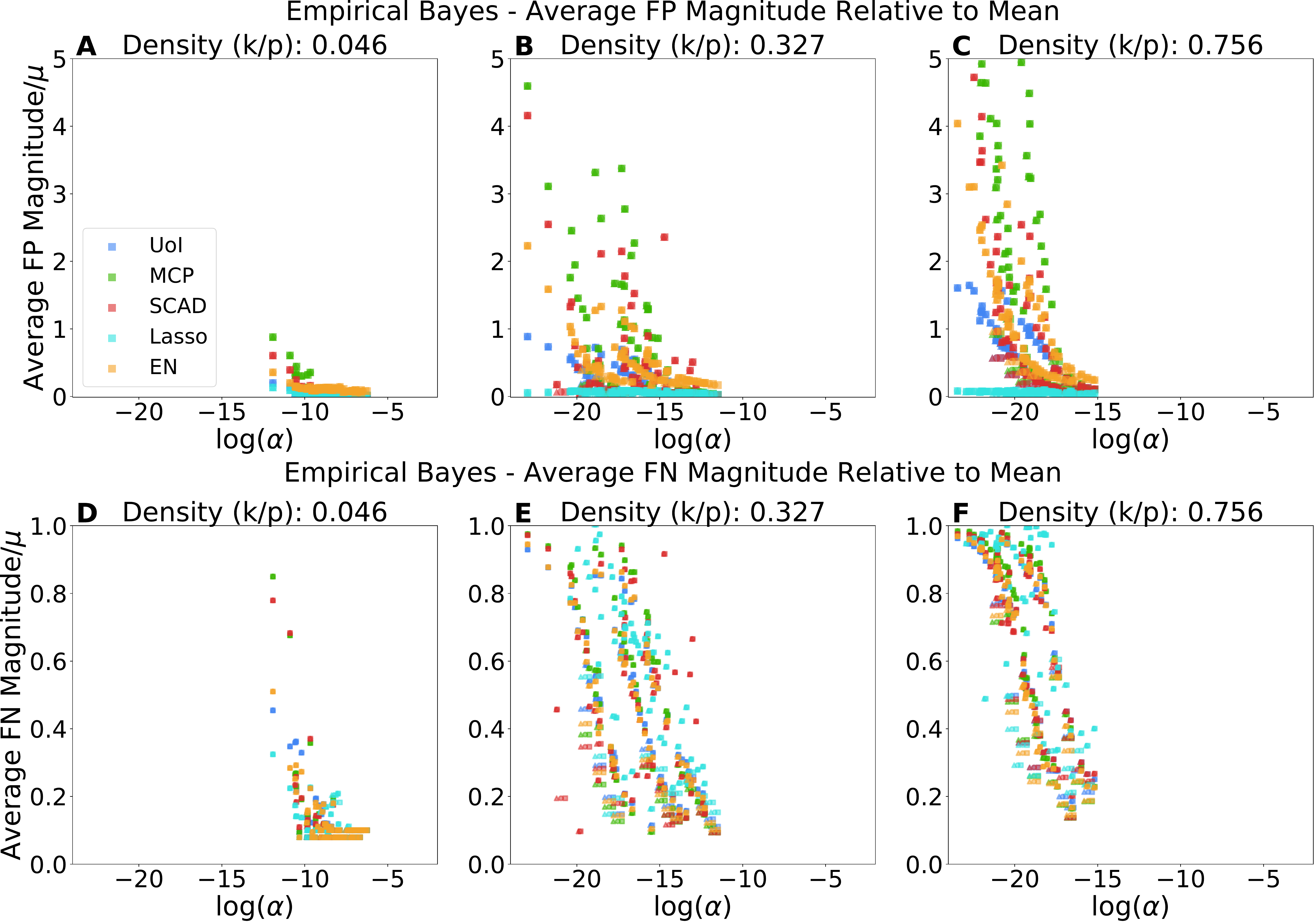}
    \captionof{figure}{Panels A-C: Average magnitude of false positives divided by the mean of the non-zero $\beta$ coefficients for empirical Bayes selection. Panels D-F: Average magnitude of false negatives divided by the mean of the non-zero $\beta$ coefficients.}
    \label{s:eb_fnfp}
    \includegraphics[width=\textwidth]{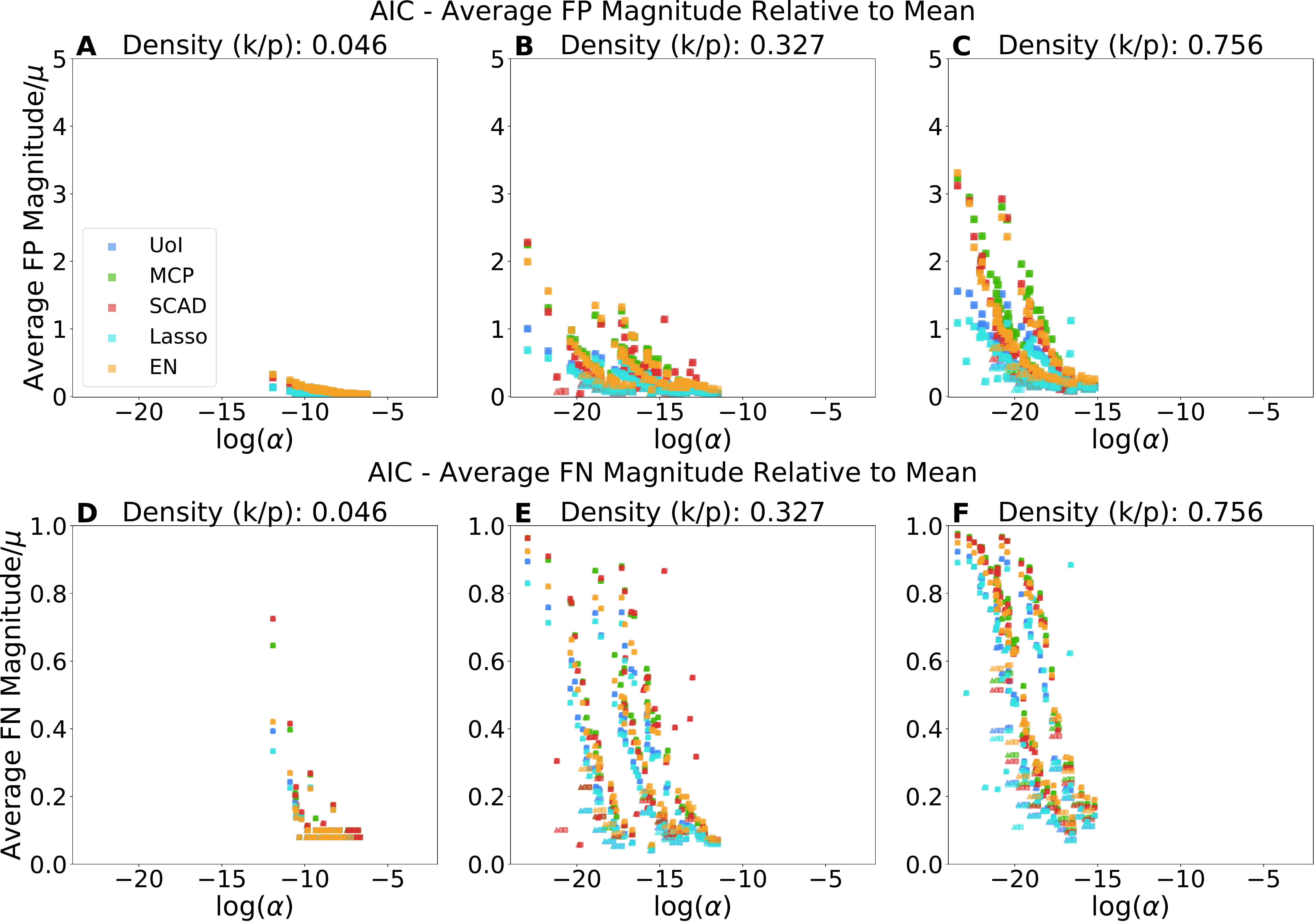}
    \captionof{figure}{Panels A-C: Average magnitude of false positives divided by the mean of the non-zero $\beta$ coefficients for AIC selection. Panels D-F: Average magnitude of false negatives divided by the mean of the non-zero $\beta$ coefficients.}    
    \label{s:aic_fnfp}
    \includegraphics[width=\textwidth]{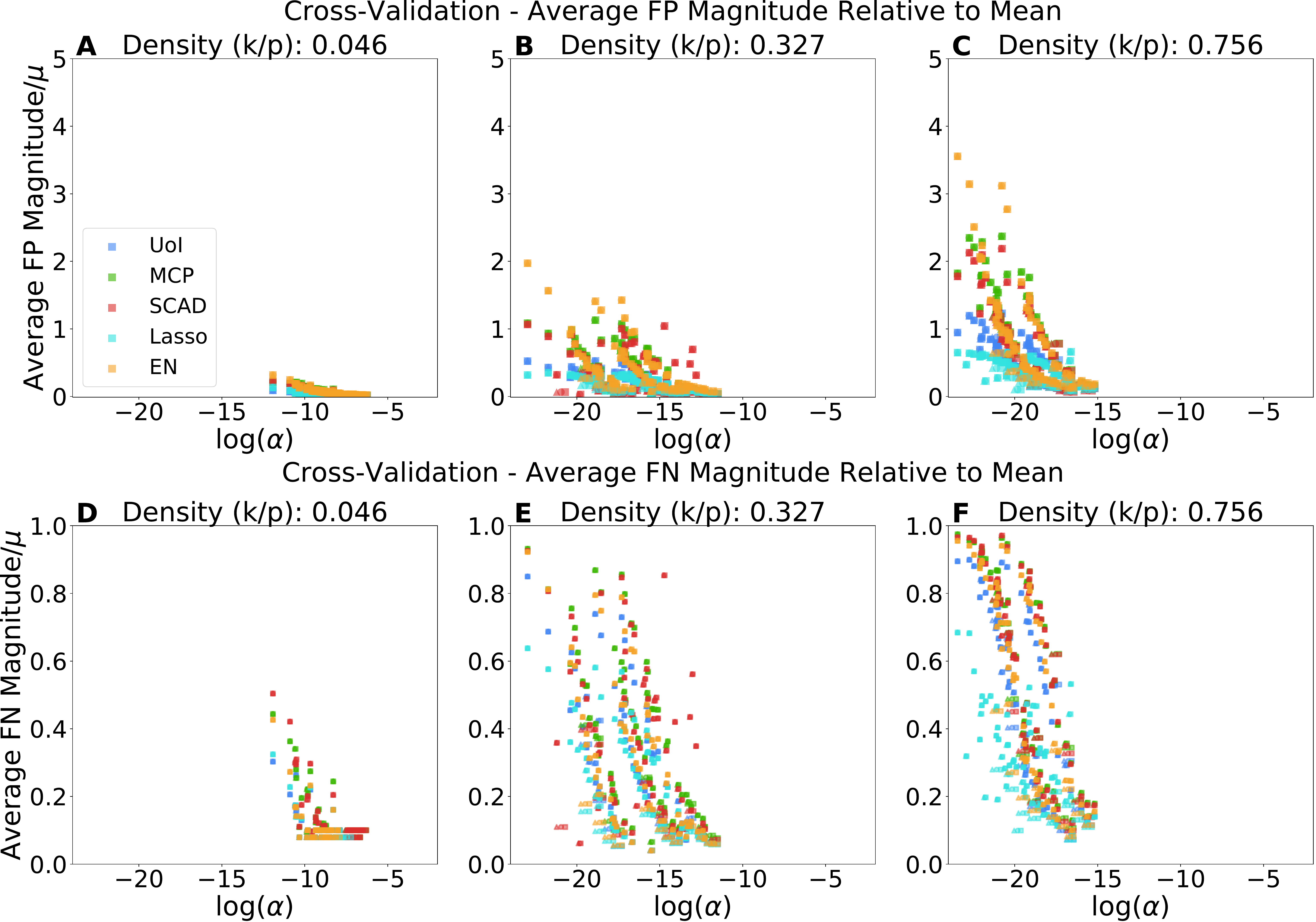}
    \captionof{figure}{Panels A-C: Average magnitude of false positives divided by the mean of the non-zero $\beta$ coefficients for cross-validation selection. Panels D-F: Average magnitude of false negatives divided by the mean of the non-zero $\beta$ coefficients.}    
    \label{s:cv_fnfp}
\end{center}

\section{Comparison of Bias/Variance of UoI vs. SCAD/MCP}

The $UoI_{\text{Lasso}}$, SCAD, and MCP estimators, especially when combined with BIC or empirical Bayes model selection achieve state of the art model selection performance in the presence of correlated variability (Tables 2-7). The UoI algorithm separates estimation and selection by fitting OLS models to non-zero support coefficients, and uses bootstrapped aggregation to average together several model estimates. Here, we demonstrate that in addition to achieving selection accuracies comparable to SCAD/MCP for correlated designs, these features of the UoI algorithm reduced the bais and variance of estimates relative to SCAD/MCP.

In Figures \ref{s:bias_comp} and \ref{s:variance_comp} we compare the bias and variance, respectively, between UoI/MCP/SCAD for the BIC and empirical Bayes model selection criteria. The bias ($\mathbb{E}(\hat{\beta}) - \beta$, where $\hat{\beta}$ are the estimated coefficients) and variance ($\mathbb{E}(\hat{\beta} - \mathbb{E}(\beta))^2$), was estimated by averaging over 20 fit repetitions. With respect to bias, when using BIC selection, all 3 algorithms achieved essentially the same performance (panels A-C of \ref{s:aic_fnfp}). However, UoI reduced the bias over SCAD and MCP when using empirical Bayesian model selection. Curiously, SCAD and MCP featured very high bias even at large $\log \alpha$ (panel D) in sparse models, but the advantage of UoI persisted at all model densities (blue scatter points lie below the red and green scatter points). The variance of UoI was consistently lower than SCAD/MCP for both BIC and empirical Bayes, across model densities and $\log \alpha$ values. These results highlight the ability of model averaging and re-estimation procedures to reduce estimation bias and variance.

\begin{center}
    \includegraphics[width=\textwidth]{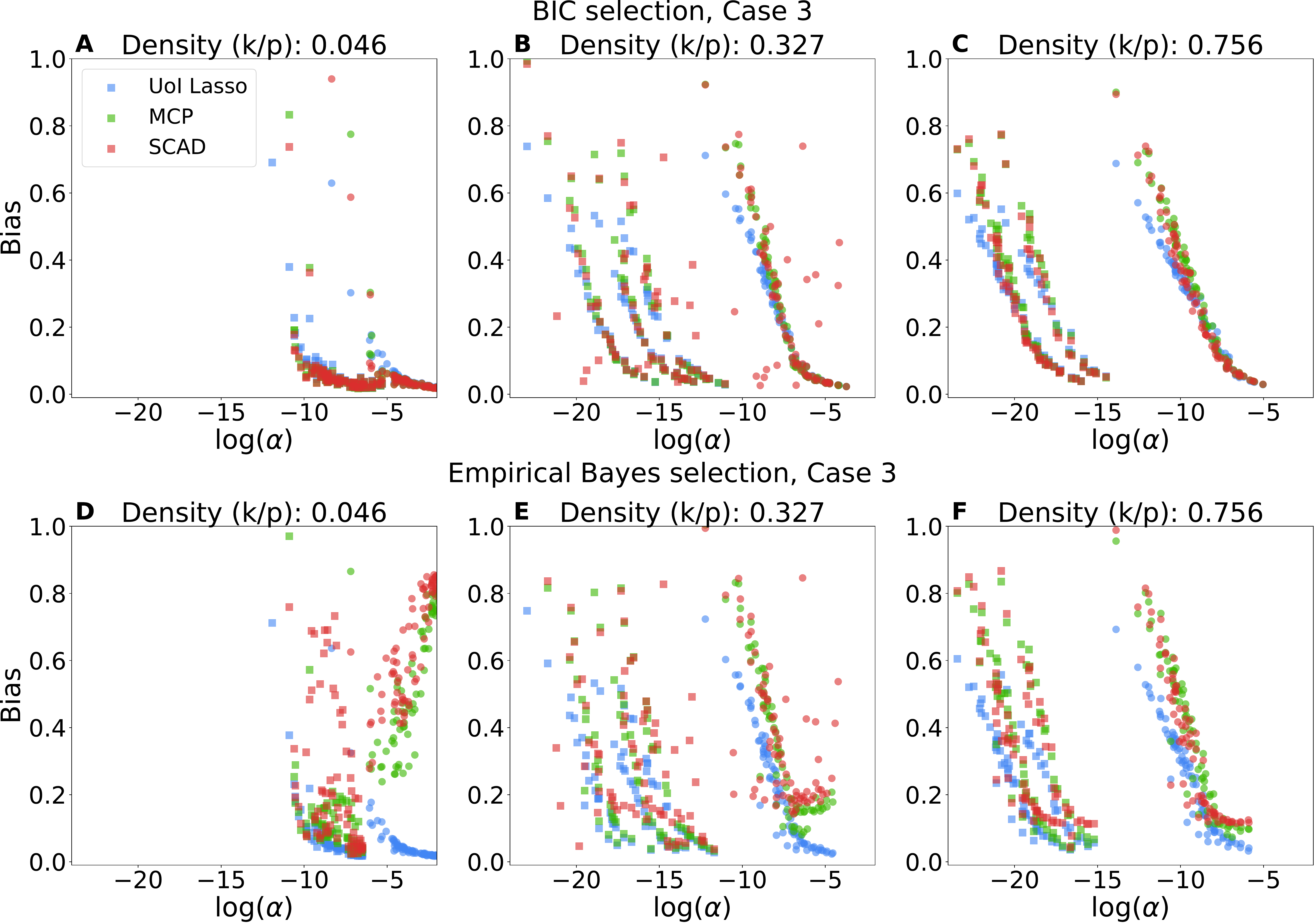}
    \captionof{figure}{Plot of Bias ($\mathbb{E}[\hat{\beta}] - \beta$) vs. $\log \alpha$ for the BIC model selection (A-C) and the empirical Bayes model selection (D-F) for Case 3 signal conditions (n/p ratio 16, SNR 10).}
    \label{s:bias_comp}
    \includegraphics[width=\textwidth]{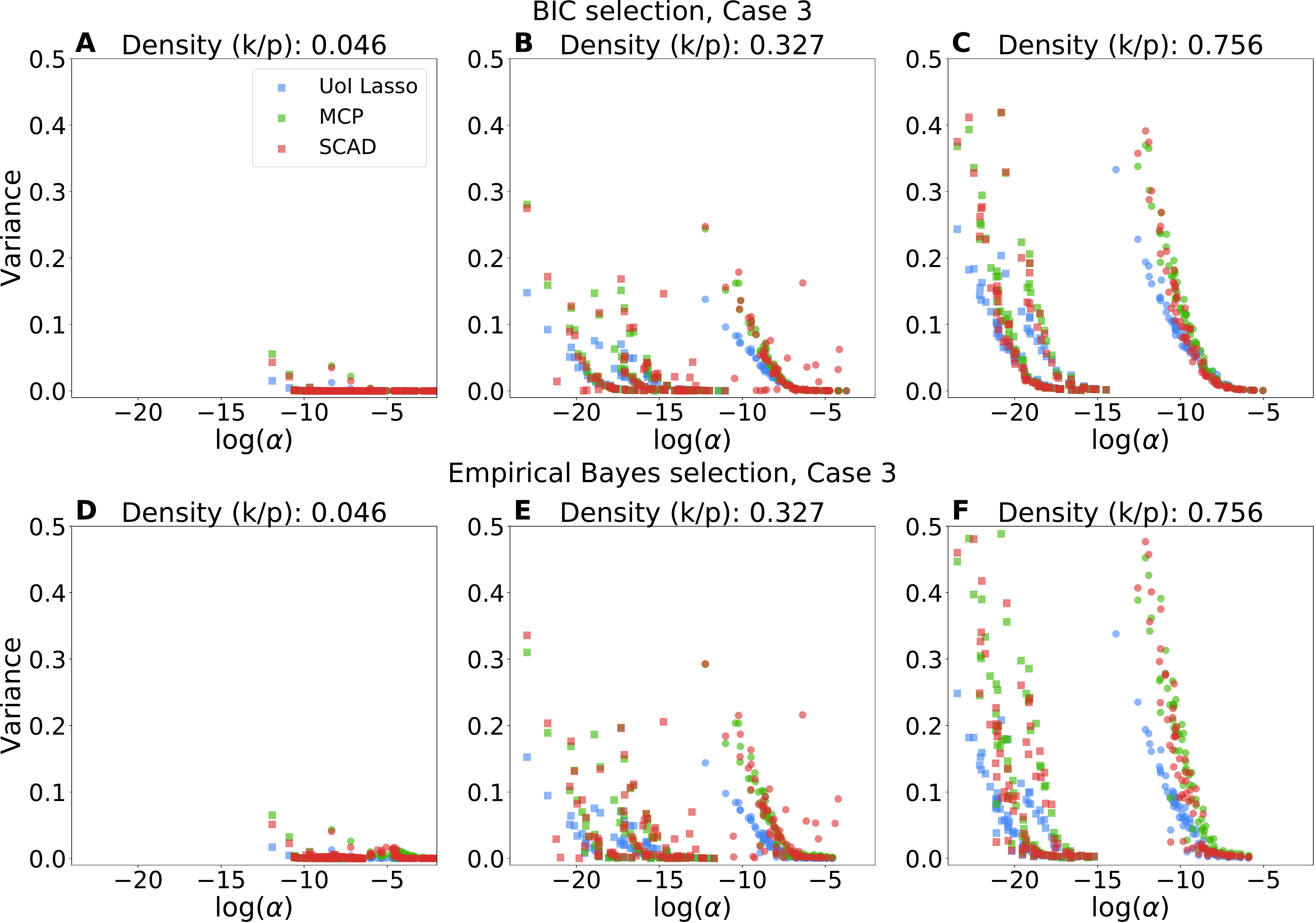}
    \captionof{figure}{Plot of Variance ($\mathbb{E}(\hat{\beta} - \mathbb{E}(\beta))^2$) vs. $\log \alpha$ for the BIC model selection (A-C) and the empirical Bayes model selection (D-F) for Case 3 signal conditions (n/p ratio 16, SNR 10).}
    \label{s:variance_comp}
\end{center}

\section{Comparison with the Irrepresentable Constant}

In \cite{zhao_model_2006}, the importance of the irrepresentable constant in ensuring the (asymptotic) selection consistency of the Lasso was established. Specifically, if we let $S \in \mathcal{I}_k$ index the true model support and let $\bar{S} := \{1, ..., p \} \setminus S$ index the complement of the model support, we can partition the feature covariance matrix as follows:

$$\Sigma = \begin{bmatrix} \Sigma_{SS} & \Sigma_{S,\bar{S}} \\ \Sigma_{\bar{S}, S} & \Sigma_{\bar{S} \bar{S}} \end{bmatrix}$$

Letting $\beta_S$ denote the vector of non-zero coefficients. The irrepresentable constant (section 3.2 in \cite{zhao_model_2006}) is then given by $\eta = 1 - |\Sigma_{\bar{S}, S}\Sigma_{SS}^{-1} \; \text{sign}(\beta_S)|_\infty$. For $\eta < 0$, the Lasso is not asymptotically selection consistent. 

An example of how $\eta$ tunes the finite sample selection accuracy of the Lasso and the other estimators considered, we calculated $\eta$ for the design matrices considered in this study and plot the selection accuracy vs. $\eta$ for BIC selection and the Gaussian coefficient distribution in Figure \ref{saveta}. For low model densities, the relationship between $\eta$ and selection accuracy was as expected - selection accuracy of algorithms montonically decays as $\eta \to 0$. This decline was more gradual for the Lasso and Elastic Net (cyan and orange scatters), while it is quite dramatic for UoI/SCAD/MCP. As the model density increases to 0.25 and above, the model selection performance declined as $\eta \to 0$ from the right, but rebounds for $\eta < 0$. At model density 0.25 (panel C), this effect was especially pronounced for the Lasso and Elastic Net. As the model density increases, a higher proportion of the feature covariance matrices considered in this study corresponded to $\eta < 0$, while the selection accuracy was no longer monotonically related to $\eta$. In panels D-F, one observes that the selection accuracy declined as $\eta \to 0$ from both the left and right, but that the selection accuracy was only slightly reduced from its maximum for the most negative values of $\eta$. This observation holds for all estimators.

To explain this observation, we plot in Figure \ref{rhovseta} a scatter of the irrepresentable constant vs. $\rho(\Sigma, k)$ for the same model densities as in Figure \ref{saveta}. In panel A, a monotonic relationship between $\eta$ and $\rho(\Sigma, k)$ is observed, with $\rho(\Sigma, k) \to 0$ as $\eta \to 0$. Recalling that smaller $\rho(\Sigma, k)$ represents a harder support recovery problem, the two quantities tracking each other thus matches with the selection accuracy performance observed in Figure \ref{rhovseta}(A). Beginning saliently in panel \ref{rhovseta}(C) and continuing in panels D-F, we observe that design matrices with $\eta < 0$ actually yielded relatively large $\rho(\Sigma, k)$, with small $\rho(\Sigma, k)$ corresponding to matrices with the smallest $|\eta|$. As this pattern mirrors that of the selection accuracy observed in Figure \ref{saveta}, we conclude that $\eta$ tracks the \emph{finite sample} selection accuracy performance of the Lasso (and to a lesser extent other estimators) only insofar as it is monotonically related to $\rho(\Sigma, k)$. In other words, $\rho(\Sigma, k)$ is a more reliable 
measure of how feature covariance matrices modulate selection accuracy. Note that in \cite{zhao_model_2006}, empirical evaluation was done on the probability that the entire Lasso solution path would contain the true support, \emph{not} on the selection accuracy after employing a model selection criteria.

\begin{center}
\includegraphics[width=\textwidth]{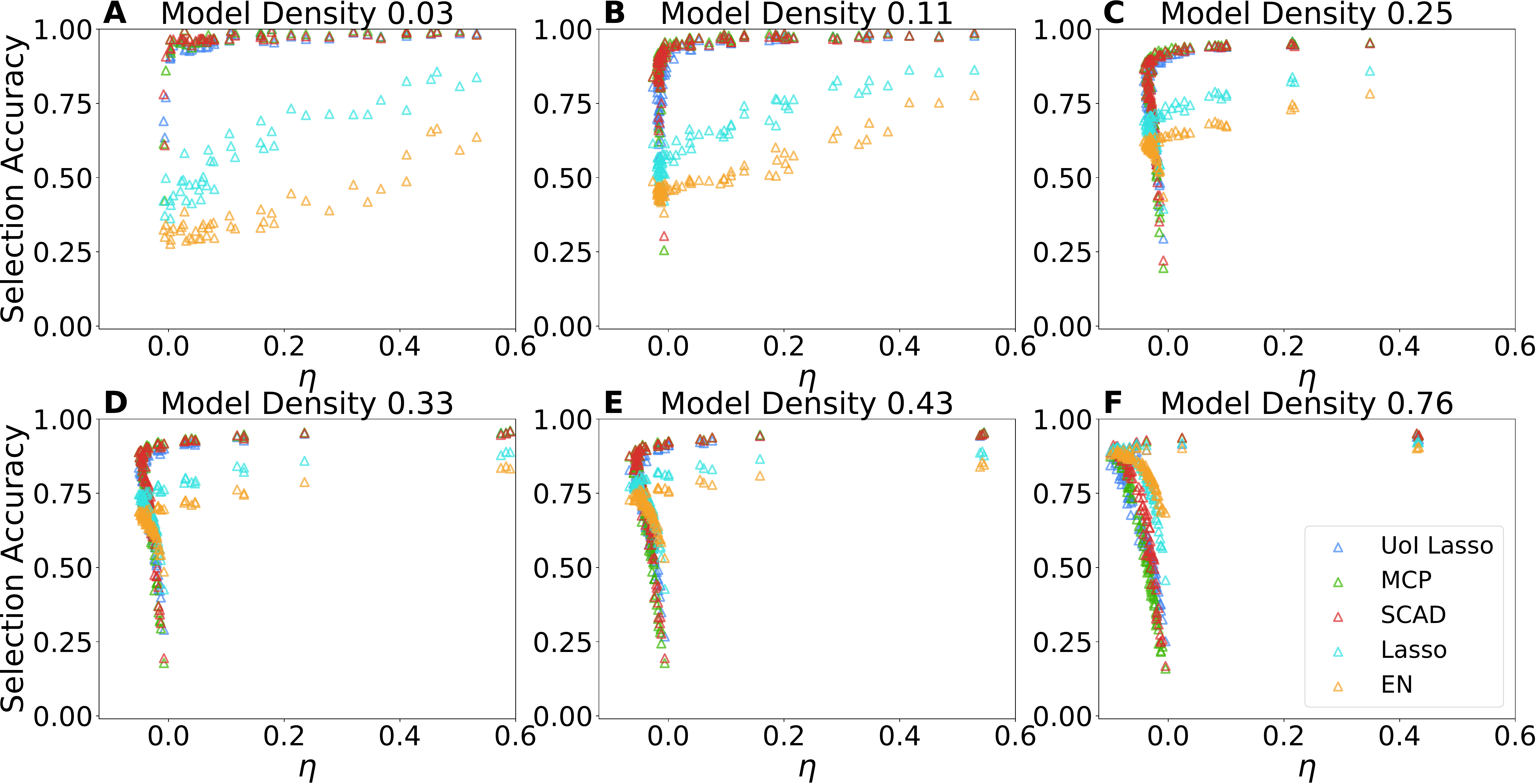}
\captionof{figure}{Plot of selection accuracy vs. $\eta$ for BIC selection criteria and inverse exponential $\beta$ distribution for different model densities. At low model densities, the decay in selection performance is monotonic as $\eta \to 0$, whereas for higher model densities, the selection accuracy decays rapidly with $|\eta|$, but selection accuracies for regression problems arising from design matrices correspond to $\eta < 0$ are high.}
\label{saveta}
\end{center}

\begin{center}
    \includegraphics[width=\textwidth]{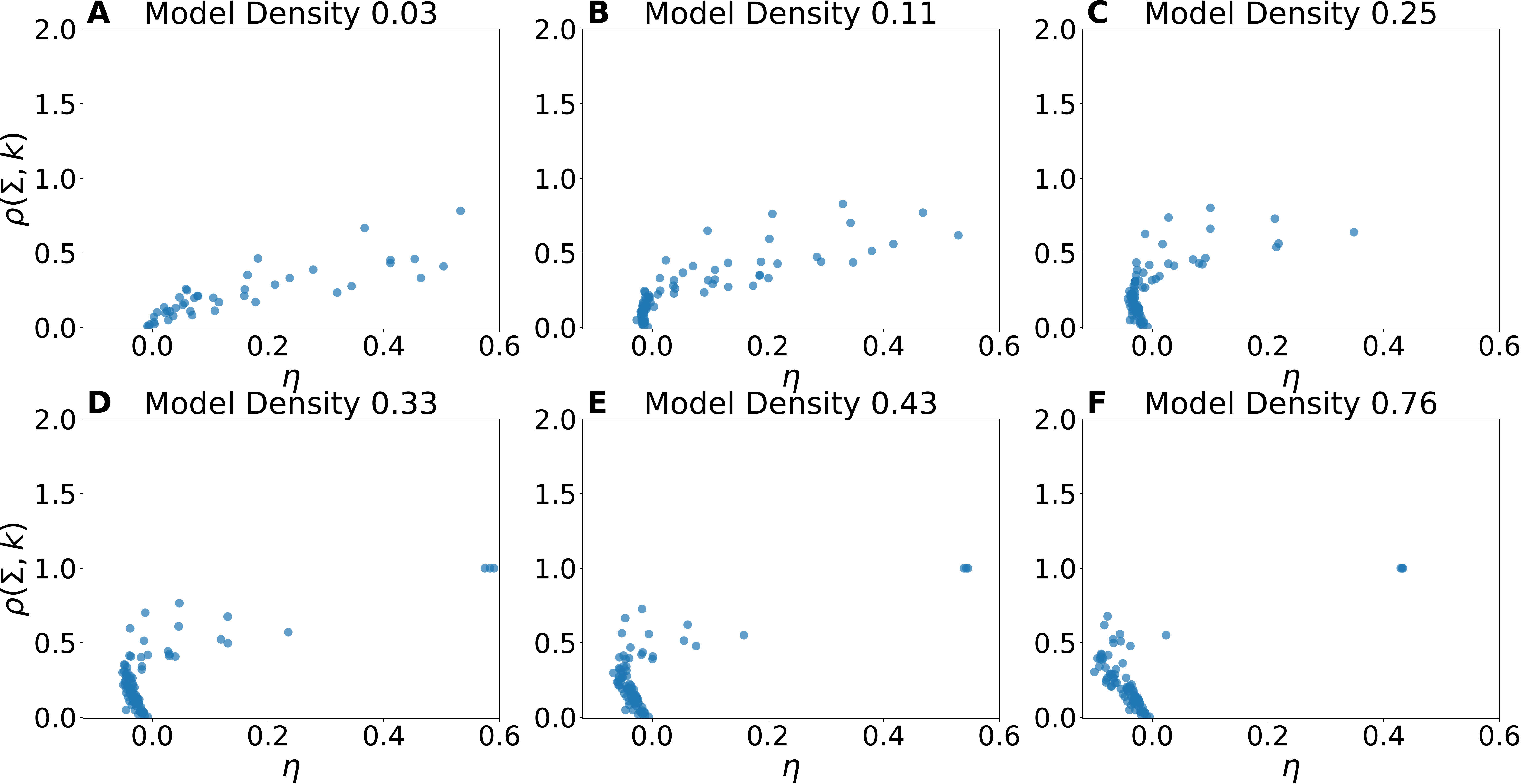}
    \captionof{figure}{Plot of $\rho(\Sigma, k)$ vs. $\eta$ across model densities. Note that $k = [\text{Model Density} \times 500]$. The profile of scatter points resembles that of the selection accuracies in Figure \ref{saveta}}
    \label{rhovseta}
\end{center}

\end{document}